\documentclass[aps,twocolumn,10pt,pra,superscriptaddress,amsmath,amssymb]{revtex4-1}

\usepackage{amssymb,amsfonts,amsmath}
\usepackage{graphicx}
\usepackage{color}
\usepackage{txfonts}
\usepackage[T1]{fontenc}
\usepackage[utf8]{inputenc}
\usepackage{CJK}
\usepackage{comment}
\usepackage{setspace}
\usepackage[usenames,dvipsnames]{xcolor}
\usepackage{soul}
\usepackage[normalem]{ulem}
\usepackage{listings}
\usepackage{hyperref}
\usepackage{type1cm}
\usepackage{lettrine}
\usepackage{courier}

\definecolor{dkgreen}{rgb}{0,0.6,0}
\definecolor{gray}{rgb}{0.5,0.5,0.5}
\definecolor{mauve}{rgb}{0.58,0,0.82}

\newcommand{\jokbo}{\textit{jokbo}}

\begin{document}
\begin{CJK}{UTF8}{mj} 

\title{Matchmaker, Matchmaker, Make Me a Match: Migration of Populations via Marriages in the Past}

\author{Sang Hoon Lee (이상훈)}
\thanks{These authors contributed equally to this work.}
\affiliation{Integrated Energy Center for Fostering Global Creative Researcher (BK 21 plus) and Department of Energy Science, 
Sungkyunkwan University, Suwon 440--746, Korea}
\affiliation{Oxford Centre for Industrial and Applied Mathematics (OCIAM), Mathematical Institute, University of Oxford, 
Oxford, OX2 6GG, United Kingdom}

\author{Robyn Ffrancon}
\thanks{These authors contributed equally to this work.}
\affiliation{Department of Physics, University of Gothenburg, 412 96
Gothenburg, Sweden}

\author{Daniel M. Abrams}
\affiliation{Department of Engineering Sciences and Applied Mathematics, Northwestern University, Evanston, Illinois 60208, USA}

\author{Beom Jun Kim (김범준)}
\affiliation{Department of Physics, Sungkyunkwan University, Suwon, 440-746, Korea}

\author{Mason A. Porter}
\affiliation{Oxford Centre for Industrial and Applied Mathematics (OCIAM), Mathematical Institute, University of Oxford, 
Oxford, OX2 6GG, United Kingdom}
\affiliation{CABDyN Complexity Centre, University of Oxford, Oxford, OX1 1HP, United Kingdom}

\begin{abstract}
The study of human mobility is both of fundamental importance and of great potential value.  For example, it can be leveraged to facilitate efficient city planning and improve prevention strategies when faced with epidemics.  The newfound wealth of rich sources of data---including banknote flows, mobile phone records, and transportation data---has led to an explosion of attempts to characterize modern human mobility.  Unfortunately, the dearth of comparable historical data makes it much more difficult to study human mobility patterns from the past. In this paper, we present an analysis of long-term human migration, which is important for processes such as urbanization and the spread of ideas. We demonstrate that the data record from Korean family books (called ``\jokbo{}'') can be used to estimate migration patterns via marriages from the past 750 years. We apply two generative models of long-term human mobility to quantify the relevance of geographical information to human marriage records in the data, and we find that the wide variety in the geographical distributions of the clans poses interesting challenges for the direct application of these models. Using the different geographical distributions of clans, we quantify the ``ergodicity'' of clans in terms of how widely and uniformly they have spread across Korea, and we compare these results to those obtained using surname data from the Czech Republic. To examine population flow in more detail, we also construct and examine a population-flow network between regions. Based on the correlation between ergodicity and migration in Korea, we identify two different types of migration patterns: diffusive and convective.  We expect the analysis of diffusive versus convective effects in population flows to be widely applicable to the study of mobility and migration patterns across different cultures.

\end{abstract}




\maketitle


\section{introduction}

Since Quetelet's advocacy of ``social physics'' in the 1830s~\cite{Quetelet1835} and Ravenstein's seminal work later in the nineteenth century~\cite{Ravenstein1885}, quantitative studies of human mobility have suggested that human movements follow statistically predictable patterns~\cite{Stouffer1940, orig_grav_mobil_pap, Wilson1967, Thornthwaite1934, Lucas1981, Dorigo1983, Greenwood2003, Stewart1950}.  Such systems-level studies are an important complement to individual-based approaches, as they can reveal population-level phenomena that are difficult to deduce by focusing on the characteristics of isolated members
\cite{cohen-pnas2008}.

Research that takes a physics-based approach has focused predominantly on modern mobility---rather than historical mobility and migration---because of the disproportionate availability of large, rich data sets from modern life~\cite{underst_hum_mobil_patt, scalling_laws_trav, online_game_mobil, Davies2013, Asgari2013}.  By contrast, historical data tend to be sparse, incomplete, and noisy. These constraints limit the scope of conclusions that one can draw about how humans mingled, mixed, and migrated over long time scales~\cite{Park1928, Sjaastad1962}. In this paper, we investigate {historical} human mobility and associated human migration by studying the matchmaking process for traditional marriages in Korea combined with modern census data in South Korea.  We obtain our data from Korean ``family books''  called \jokbo{} (족보 in Korean). Such a confluence of historical and modern data is rare, and it allows a novel test of generative models for human mobility.

According to Korean tradition, family names are subdivided into clans called \textit{bon-gwan} (본관), which are identified by a unique place of origin.  For example, the two Korean authors of this paper belong to the clans ``Kim from Gimhae (김해 김)'' and ``Lee from Hakseong (학성 이)'', and the clan ``Lee from Hakseong'' is distinct from the clan ``Lee from Jeonju (전주 이)'' [the royal clan of the Joseon dynasty and the Great Korean Empire (1392--1910)]. When two Koreans marry, the bride's clan and her birth year are customarily recorded in the \jokbo{} owned by the groom's family. These \jokbo{} are kept in the groom's family and passed down through the generations; they serve primarily as a record of the names and birth years of all male descendants{~\cite{JokboWikipedia,KoreanFamilyNamesAndGenealogies}}. In previous work, researchers used the marriage data contained in these books to estimate the population sizes and distributions of clans in Korea as far as 750 years in the past \cite{Kiet2007,SKBaek2007,tenthoukim}. Such distributions are useful for understanding quantitative aspects of human culture, and we proceed even further by conducting a systematic investigation of the geographical information embedded in \jokbo{}.

We examine a set of ten \jokbo{} to try to understand how geographical separation affected human interaction in the past in Korea.  Specifically, we examine how interclan marriage rates can be predicted by physical distance and how clans themselves have spread across the country during the past several hundred years.  To do this, we apply two generative models for describing human mobility patterns to \jokbo{} records of past marriages between two clans.  Note that the identification of clans with specific geographical origins is not unique to Korea. For example, the origins of British and Czech surnames were also the subject of recent investigations~\cite{UK_surname,czech_surname_pap}.  

Our analysis consists of two parallel approaches. First, we use marriages recorded in \jokbo{} to obtain snapshots of migration (mainly of individual women) for a ``marriage-flux analysis''.  We apply two generative models for population flow, discuss the results of applying these models, and explain the limitations that arise from the wide variety in the geographical-distribution patterns of the clans.  Second, to consider the geographical spread of clans in more detail, we conduct an ``ergodicity analysis''. We use the modern geographical distribution of clans from census data to infer ``ergodicity'' of clans (mainly caused by past movement of male descent lines).  To provide an additional perspective, we also use these data to construct a network model of population flows. To the best of our knowledge, the notion of diffusive versus convective population flow is new for data-driven studies of human mobility and migration, and we believe that this kind of approach can provide valuable insights for many problems in population mobility and migration. In the present paper, we focus on long-term migration, which has significant effects on many processes over a variety of spatial and temporal scales. Such processes include urban population growth and the demographic structure of cities \cite{UN2014}; city infrastructure and planning \cite{Batty2008}; unemployment \cite{Todaro1969}; and the spread of culture, religion, and other ideas \cite{Hoerder2002}. Most early studies of migration emphasized so-called ``internal migration'' (i.e., movement within a country) \cite{Ravenstein1885,Stouffer1940,Greenwood1997,Lucas1997} like the phenomena that we investigate, though international migration is also a prominent field of study~\cite{Lee1966,Cushing2003}.  It has been more popular to study international migration than internal migration during the past few decades, but present-day urbanization processes in Asia, Africa, and Latin America have led to renewed interest in internal migration~\cite{UN2014,Lucas1997}.  We hope that our work provides useful ideas to help solve some of the fundamental questions in the migration literature: who migrates, why people migrate, and the consequences of migration (e.g., rural depopulation).

The remainder of our paper is organized as follows. In Sec.~\ref{sec:data_sets}, we introduce the \jokbo{} and census data that we use in our investigation. In Sec.~\ref{sec:methods}, we present our primary methodology for data analysis: the gravity and radiation models for marriage-flux analysis, a special case of the gravity model that we call the population-product model, and a diffusion model for ergodicity analysis. We present our main results in Sec.~\ref{sec:results}, and we conclude in Sec.~\ref{sec:conclusions}. We include detailed information on the data sets, data cleaning, additional results and practical considerations for our analysis, an investigation of a network model for population flow, and various other results in Appendices~\ref{A:jokbo_data}--\ref{sec:other_results}.

\section{Data Sets}
\label{sec:data_sets}

\subsection{\emph{Jokbo} data sets}
For our marriage-flux analysis, we use the same ten \jokbo{} data sets that were employed in \cite{Kiet2007,SKBaek2007,tenthoukim}.  An individual book contains between $1\,873$ and $104\,356$ marriage entries, and there are a total of $221\,598$ entries across all books. (See Table~\ref{table_num_ent} and Figs.~\ref{num_ent_num_fam_fig} and \ref{collaps_num_ent_year_linear} in Appendix~\ref{A:jokbo_data} for details.)  Each entry contains the bride's clan and year of birth~\cite{jokbo_details}. The oldest book has entries that date back to the 13th century. 

\begin{table*}[t]
\caption{Number of entries and other information available in each \jokbo{}, values that we determined by using additional data that we obtained from other sources, and a summary of some of our computational results for the clan corresponding to each \jokbo{}.  For each \jokbo{}, we indicate the ID (1--10), the year $t_0$ of its earliest entry, its number of entries $N_e$, and the number of distinct clans (including at least one bride for each clan) $N_c$ among those entries \cite{Kiet2007}. 
The quantity $N_{\gamma=0}$ gives the number of clans from the 2000 census (which is 4\,303) plus the number of clans in each \jokbo{} that are not already in the census. We can use these $N_{\gamma = 0}$ clans in the gravity model when $\gamma = 0$ (i.e., for the population-product model, which is applicable without geographical information) and $\alpha = 1$. (See the discussion in Appendix~\ref{sec:census}.) 
We also indicate the best values for the fitting parameters $\alpha$ and $\gamma$ of the gravity model in Eq.~\eqref{grav_mod_eq}.  We apply this fit to the brides' side of marriages, and we calculate these values by  minimizing the sum of squared differences using the \texttt{scipy.optimize} package in {\sc Python}~\cite{ASciPyOptimization} (with initial values of $\alpha = \gamma = a_G = 1.0$ in our computations).
We compute the number of administrative regions $N_{\mathrm{admin}}$ in which the clan that corresponds to each \jokbo{} (i.e., the grooms' side) appears based on census data from 1985 and 2000. We use the census data to compute a radius of gyration $r_g$ (km) for both 1985 and 2000 and to estimate a diffusion constant $D$ (km$^2$/year) for diffusion of clans between those two years. We consider clans with $N_\textrm{admin}^\textrm{2000} \geq 150$ to be ergodic (see Fig.~\ref{erg_vs_dist_fig}).  Based on this definition, all ten clans in the \jokbo{} data are ergodic.
}
\begin{ruledtabular}
\begin{tabular}{rrrrrrrrrrrcr}
ID & $t_0$ & $N_e$ & $N_c$ & $N_{\gamma=0}$ & $\alpha$ & $\gamma$ & $N_\textrm{admin}^\textrm{1985}$ & $N_\textrm{admin}^\textrm{2000}$ & $r_g$ (1985) & $r_g$ (2000) & Ergodic? & $D$ \\
\hline
   1 & 1513  & 104\,356 & 2\,657 & 5\,510 & $1.0749$ & $-0.0349$ & 199 & 199 & 115.5 & 113.5 & Y & 0.062  \\
   2 & 1562  &  29\,139 & 1\,274 & 4\,796 & $1.0145$ & $0.2305$  & 199 & 199 & 124.4 & 128.7 & Y & 0.737  \\
   3 & 1752  &   3\,500 & 390    & 4\,364 & $1.0853$ & $0.2000$  & 199 & 199 & 132.7 & 151.5 & Y & 0.426  \\
   4 & 1698  &  15\,445 & 915    & 4\,524 & $0.9678$ & $0.1210$  & 199 & 199 & 132.7 & 151.5 & Y & 0.426  \\
   5 & 1439  &  17\,911 & 923    & 4\,551 & $0.9452$ & $0.2346$  & 198 & 199 & 101.2 & 97.4  & Y & 0.062  \\
   6 & 1476  &  16\,379 & 727    & 4\,462 & $1.1102$ & $0.5377$  & 130 & 196 & 144.6 & 128.8 & Y & 2.253  \\
   7 & 1802  &   1\,873 & 289    & 4\,359 & $1.4930$ & $-0.0961$ & 199 & 199 & 110.2 & 116.1 & Y & $-0.062$ \\
   8 & 1254  &  15\,006 & 958    & 4\,570 & $0.9651$ & $0.1285$  & 198 & 198 & 114.1 & 109.6 & Y & 0.101  \\
   9 & 1458  &   6\,463 & 548    & 4\,376 & $1.1253$ & $0.3650$  & 196 & 195 & 118.6 & 121.5 & Y & 0.784  \\
   10 & 1475 &  11\,526 & 736    & 4\,463 & $0.9947$ & $0.4502$  & 198 & 196 & 117.7 & 127.7 & Y & 0.461  \\
\end{tabular}
\end{ruledtabular}
\label{table_num_ent} 
\end{table*}

Previous studies of this data set \cite{Kiet2007,SKBaek2007,tenthoukim} did not use any of the information that is encoded implicitly in the geographical origins of each clan.  Such information, together with the modern geographical distribution of clans, comprises a key ingredient of our analysis.  We convert location names to geographical coordinates using the Google Maps Application Programming Interface (GMAPI)~\cite{google_maps}. Because of the much sparser coverage of North Korean regions by Google Maps (see Fig.~\ref{origin_distribution} in Appendix~\ref{sec:geographical_information_from_google_maps}), this geolocation data are a biased sample of the full data. However, data for the southern half of the Korean peninsula are rich~\cite{southKoreaabund}, and it is sufficient to draw interesting and robust conclusions. For example, the effect of a change in the legality of intraclan marriage in 1997 is clearly observable in the data.

\subsection{Modern name distributions}

In addition to the \jokbo{} data sets that we employ for marriage-flux analysis, we also use data from two Korean census reports (1985 and 2000) to evaluate the current spatial distribution of clans in Korea~\cite{population_census,regional_boundary}.  As illustrated in Fig.~\ref{GimhaeKim_vs_HakseongLee}, some clans have dispersed rather broadly but others remain localized (usually near their place of origin).  Drawing on ideas from statistical mechanics \cite{ergodicity_stat_book, ergodicity_hist_stat}, we use the term ``ergodic'' as an analogy to describe clans that have spread broadly throughout Korea. We suppose that such clans have reached a dynamic equilibrium: An ergodic clan is ``spread equally'' throughout Korea in the sense that one expects it to have roughly the same geographical distribution as the population as a whole.  Note that we do not expect an ergodic clan to reach a spatially uniform state for the same reasons that the full population is not spatially uniform (e.g., inhomogeneities in natural resources, advantages to congregating in cities, etc.). 

Nonergodic clans should have rather different distributions from those that we dub ergodic because their distribution must differ significantly from that of the full population.  One can construe the notion of ergodicity as a natural extension of other physical analogies that were used in previous quantitative studies (including the original ones) on human migration \cite{Quetelet1835, Ravenstein1885, Stouffer1940, orig_grav_mobil_pap, Wilson1967, Thornthwaite1934, Lucas1981, Dorigo1983, Greenwood2003, Stewart1950}. As we discuss later, we can quantify the extent of clan ergodicity.

\section{Methods}
\label{sec:methods}

\subsection{Generative models for marriage-flux analysis}
\label{sec:generative_models}

We compute a ``marriage flux''---the rate of marriage of women from clan $i$ into clan $j$---for all clan pairs $(i,j)$ in our data~\cite{ten_possibilities}. Historically, professional matchmakers were employed to travel between families to arrange marriages~\cite{book_matchmaker}, so we posit that physical distance plays a significant role in determining marriage flux.  We examine this hypothesis using two generative models: a conventional gravity model with adjustable parameters that incorporates the distance between regions and the effects (or lack thereof) of each region's population~\cite{orig_grav_mobil_pap,Wilson1967}, and a recently developed, parameter-free radiation model~\cite{orig_rad_mod_pap, gravvsradmodel}.

The gravity model has been used to explain phenomena such as commuting patterns and disease spread~\cite{Young1924,highw_grav_pap, subwa_grav_pap, diseases_grav_pap}. In this model, the flux of population $G_{ij}$ from site $i$ to site $j$ is
\begin{equation}\label{grav_mod_eq}
  	G_{ij} = \frac{m^{\alpha}_i m^{\beta}_j}{r^{\gamma}_{ij}}\,,    
\end{equation}
where $\alpha$, $\beta$, and $\gamma$ are adjustable exponents. For our purposes, $G_{ij}$ is proportional to the flux of women from clan $i$ to clan $j$ through marriage. The total population of clan $i$ is given by $m_i$, and the variable $r_{ij}$ is the distance between the centroids of clans $i$ and $j$.  We employ census data from 2000 to calculate centroids using the spatial population distribution for each clan \cite{population_census}.
Importantly, note that choosing $\gamma = 0$ in the gravity model yields a special case in which flux is independent of distance. As we will see in Sec.~\ref{sec:generative_model_result},
this situation arises when large uncertainties in geographical locations (due to clan ergodicity) hinder the accuracy of estimations of distances.

\begin{figure*}[t]
\includegraphics[width=0.95\textwidth]{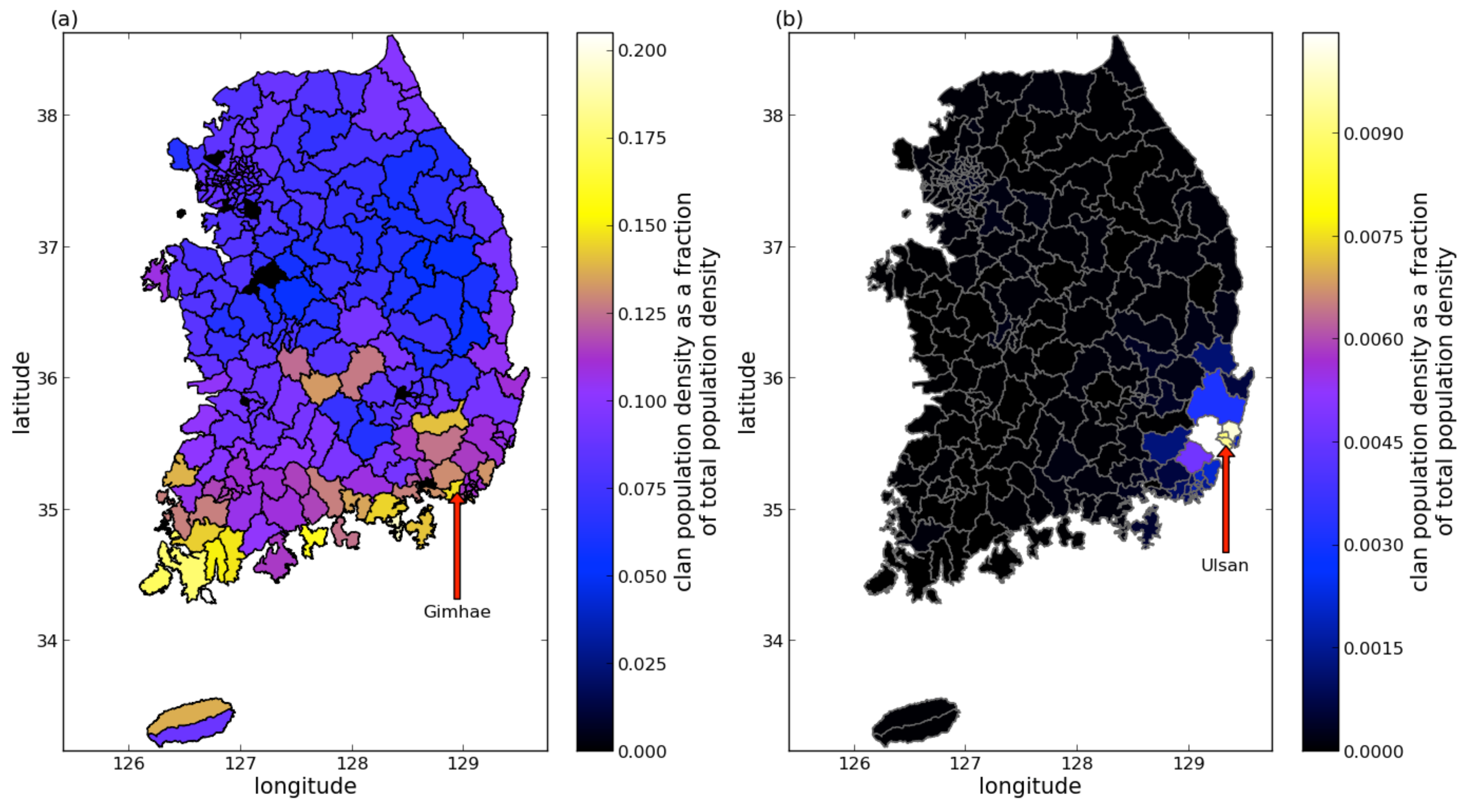}
\caption{Examples of (a) ergodic and (b) nonergodic clans. We color the regions of South Korea based on the fraction of the total population composed of members of the clan in the year 2000.  We use arrows to indicate the origins of the two clans: Gimhae on the left and Ulsan (``Hakseong'' is the old name of the city) on the right. In this map, we use the 2010 administrative boundaries~\cite{regional_boundary}. See the appendices for discussions of data sets and data cleaning.
}
\label{GimhaeKim_vs_HakseongLee}
\end{figure*}

Determining the centroid locations of clans from modern census data is more accurate than attempting to determine the locations where clans originated~\cite{origin_locations} for two reasons. First, for many clans, origin-place names 
have differed from geographical clan centers from the beginning of recorded Korean history---which, in particular, predates the period that spans our \jokbo{} data sets~\cite{KoreanFamilyNamesAndGenealogies,Donggukyeojiseungram}. Second, the origin-place names for many clans have become outdated and cannot be located accurately via the names of modern administrative regions. For instance, the clan origin ``Hakseong'' of the first author is an old name for the city Ulsan in South Korea, but the name ``Hakseong'' is currently only used to describe the small administrative region ``Hakseong-dong'' in Ulsan. However, as we demonstrate in Fig.~\ref{GimhaeKim_vs_HakseongLee}, using the centroid location of ``Lee from Hakseong'' correctly gives the modern city Ulsan. This procedure works in part because Lee from Hakseong is a nonergodic clan; for ergodic clans such as Kim from Gimhae, the spatial precision is much worse.  This is an important observation that we will discuss in detail later.

We use a version of the radiation model that takes finite-size effects into account~\cite{gravvsradmodel}. The population flux $R_{ij}$ from clan $i$ to clan $j$ is
\begin{equation}\label{rad_mod_eq}
  	R_{ij} = \frac{\Omega_i}{1 - m_i/N} \times \frac{m_i m_j}{(m_i + s_{ij})(m_i + m_j + s_{ij})}\,,
\end{equation}
where $\Omega_i=\sum_j R_{ij}$ is proportional to the total population that marries from clan $i$ into any other clan, $N$ is the total population, and $s_{ij}$ is the exclusive population within a circle of radius $r_{ij}$ centered on the centroid of clan $i$. Note that members of clans $i$ and $j$ are \textit{not} included in computing $s_{ij}$ \cite{orig_rad_mod_pap}.  As before, $m_i$ is the population of clan $i$, members of clan $i$ marry into clan $j$, and clan $j$ keeps the marriage records.
In contrast to the gravity model, the radiation model does not include any external parameters.  Importantly, this renders it unable to describe the geographically-independent situation that we need to consider in our study (and which we can obtain by setting $\gamma = 0$ in the gravity model).

For both the gravity and radiation models, we use census data from the year 2000 \cite{population_census} as a proxy for past populations.  This allows us to compute the quantities $r_{ij}$, $m_i$, and $s_{ij}$. Our approximation is supported by previously reported estimates of stability in Korean society. Historically, most clans have grown in parallel with the total population, so we assume that the relative sizes of clans have remained roughly constant \cite{tenthoukim}. In both Eqs.~\eqref{grav_mod_eq} and \eqref{rad_mod_eq}, only the relative sizes $m_i/N$ and $s_{ij}/N$ matter for calculating the flux (up to a constant of proportionality).

\subsection{Human diffusion and ergodicity analysis}
\label{sec:human_diffusion}
One way to quantify the notion of clan ergodicity is to examine what we call the ``clan-density anomaly'', which describes the local deviation in density of members of a given clan.  The clan-density anomaly is $\phi_i(\mathbf{r},t)=c_i(\mathbf{r},t)-[m_i(t)/N(t)]\rho(\mathbf{r},t)$ at position $\mathbf{r} = (x,y)$ and time $t$, where $c_i(\mathbf{r},t)$ is the (spatially and temporally varying) local clan concentration (i.e., the clan population density), $m_i(t)$ is the total clan population, $\rho(\mathbf{r},t)$ is the local population density (i.e., the total population of all clans at point $\mathbf{r}$ and time $t$, divided by the differential area), and $N(t)$ is the total population of all of the clans at time $t$.  If a clan were to occupy a constant fraction of the population everywhere in the country, then $\phi_i=0$ everywhere because its local concentration would be $c_i=(m_i/N)\rho$. (This situation corresponds to perfect ergodicity.) The range of typical values for the clan-density anomaly depends on a clan's aggregate concentration in the country.  Examining the anomaly relative to clan concentration, the year-2000 numbers for $\phi_i/(m_i \rho/N)$ range from $-1\,700$ to $7\,400$ for Kim from Gimhae and from $-19\,000$ to $87\,000$ for Lee from Hakseong.  Clearly, the distribution of the latter is much more heterogeneous (see Fig.~\ref{phi_distribution_for_Kim_and_Lee} in Appendix~\ref{sec:other_results}). 

Combining the notion of clan-density anomaly with traditional arguments---flow ideas based on Ohm's law and ``molecular weights for population'' are mentioned explicitly in \cite{Thornthwaite1934,Stewart1950}---about migration from population gradients \cite{Ravenstein1885,Stouffer1940,orig_grav_mobil_pap,Wilson1967,Thornthwaite1934,Lucas1981,Dorigo1983,Greenwood2003,Stewart1950} suggests a simple Fickian law~\cite{Thornthwaites_law} for human transport on long time scales. We propose that the flux of clan members is $\mathbf{J}_i \propto \nabla \phi_i$, so individuals move preferentially away from high concentrations of their clans.  This implies that ${\partial c_i} / {\partial t} = \nabla \cdot \mathbf{J}_i \propto \nabla^2 \phi_i$ (where we have assumed that the constant of proportionality is independent of space), which yields the diffusion equation
\begin{equation}\label{diffusion_eq}
  	\frac{\partial \phi_i}{\partial t} = D_i \, \nabla^2 \phi_i\,.
\end{equation}
We thereby identify the constant of proportionality as a mean diffusion constant $D_i$ with dimensions $[\textrm{length}^2/\textrm{time}]$. This prediction of diffusion of clan members is consistent with existing theories that posited human diffusion (e.g., cultural \cite{Ammerman1971} and demic \cite{Fort2012} diffusion).  An important distinction is that we are proposing a process of diffusive mixing of clans rather than diffusive expansion of an idea or group. If this theory is correct, then one should expect clan-density anomalies to simply diffuse over time.  One should also be able to estimate diffusion constants by comparing the spatial variation at two points in time.

One can gain insight into the above diffusion process by calculating the radius of gyration (a second moment) of the clan-density anomaly as a proxy for measuring ergodicity.  Suppose that clan $i$'s concentration $c_i(\mathbf{r},t)$ is known on a set of discrete regions $\{ S_k \}$ with areas $\{ A_k  \}$. 
We define the centroid coordinates for the $k$th region as
\begin{equation} \label{region_centroid}
	\mathbf{r}(k) = \frac{1}{\left| S_k \right|} \sum_{\mathbf{r} \in S_k} \mathbf{r}\,,
\end{equation}
where $\left| S_k \right|$ is the total number of coordinate points $\mathbf{r}$ in $S_k$ for normalization, and we henceforth use $\phi_i(k,t)$ to indicate $\phi_i[ \mathbf{r}(k),t ]$.
The centroid of the clan's anomaly has coordinates
\begin{equation}\label{C_formula}
  	\mathbf{r}_{i,\textrm{C}}(t) = \frac{1}{\phi_{i,\textrm{tot}}(t)} \sum_k \mathbf{r}(k) \phi_i(k,t) A_k \,, \\
\end{equation}
where $\mathbf{r}(k) = [x(k),y(k)]$ gives the coordinates of the centroid of region $k$ and the normalization constant is
\begin{equation}\label{normalization_constant}
	\phi_{i,\textrm{tot}} (t) = \sum_k \phi_i (k,t) A_k \,,
\end{equation}
where $\phi_i(k,t)$ is the anomaly of clan $i$ in region $k$ at time $t$. Note that we calculate the centroid of population for the $i$th clan (as opposed to the centroid of its anomaly) using analogous formulas to Eqs.~\eqref{C_formula} and \eqref{normalization_constant} in which $\phi_i$ is replaced by the concentration $c_i$. The radius of gyration (i.e., the spatial second moment) ${r_g}_i(t)$ of clan $i$ at time $t$ is then defined by
\begin{equation}\label{radius_of_gyration}
	{r_g}_i^2(t) = \frac{1}{\phi_{i,\textrm{tot}}} \sum_k \lVert \mathbf{r}(t) - \mathbf{r}_{i,C}(t) {\rVert}^2 \phi_i(k,t) A_k  \,,
\end{equation}
where $\lVert \cdot \rVert$ is the Euclidean norm.
We can use the set of radii of gyration $\{ r_g(t) \}$ from Eq.~\eqref{radius_of_gyration} as a proxy for ergodicity because (by construction) $r_{g_i}(t)$ quantifies how widely the clan-density anomaly of clan $i$ has spread across Korea~\cite{manipulation}.

We simulate Eq.~\eqref{diffusion_eq} between the known anomaly distributions from census data at $t_1 = 1985$ and $t_2 = 2000$ to estimate a best-fit diffusion constant $D_i$ for each clan.  We compare our results to a null model in which movement is diffusive but driven by the aggregate population density in each region rather than by clan-population anomaly.  Our clan-based diffusion model performs better than the null model for approximately 84\% of the clans.

\section{Results}
\label{sec:results}

\subsection{Marriage-flux analysis based on \emph{jokbo} and modern census data}
\label{sec:generative_model_result}

\begin{table*}[t]
\caption{Gravity-model parameters $\alpha$ and $\gamma$ in Eq.~\eqref{grav_mod_eq}
calculated for temporally-divided entries of \jokbo{} 1 by minimizing the sum of squared differences using the \texttt{scipy.optimize} package in {\sc Python}~\cite{ASciPyOptimization}.  (We again use initial values of $\alpha = \gamma = a_G = 1.0$ in these computations.) We sort the list of brides according to birth year, (temporally) partition the data such that each time window (except for the last one) has 10\,001 entries, and indicate the mean and median birth year in each window.
}
\begin{ruledtabular}
\begin{tabular}{rccrr}
Window & Year (mean) & Year (median) & $\alpha$ & $\gamma$ \\
\hline
$1$--$10\,001$ & 1739.72 & 1756 & $1.0943$ & $-0.1019$ \\
$10\,002$--$20\,002$ & 1828.51 & 1829 & $1.1130$ & $-0.0396$ \\
$20\,003$--$30\,003$ & 1865.08 & 1865 & $1.1186$ & $-0.0776$ \\
$30\,004$--$40\,004$ & 1890.72 & 1891 & $1.1277$ & $-0.0272$ \\
$40\,005$--$50\,005$ & 1910.91 & 1911 & $1.0802$ & $0.0209$ \\
$50\,006$--$60\,006$ & 1926.80 & 1927 & $1.0463$ & $0.0270$ \\
$60\,007$--$70\,007$ & 1938.99 & 1939 & $1.0886$ & $-0.0146$ \\
$70\,008$--$80\,008$ & 1949.64 & 1950 & $1.0405$ & $0.0027$ \\
$80\,009$--$90\,009$ & 1958.01 & 1958 & $1.0443$ & $-0.0807$ \\
$90\,010$--$100\,010$ & 1964.90 & 1965 & $1.0030$ & $-0.0247$ \\
$100\,011$--$104\,356$ & 1971.78 & 1971 & $1.0240$ & $-0.1077$ \\
\end{tabular}
\end{ruledtabular}
\label{jokbo1_temporally_divided} 
\end{table*}

We apply a least-squares fit on a doubly logarithmic scale to determine the coefficients $\alpha$ and $\gamma$ from Eq.~\eqref{grav_mod_eq} (along with the proportionality coefficient $a_G$, which is essentially a normalization constant, for the total number of marriages). The parameter $\beta$ is irrelevant for the aggregated entries in a single \jokbo{} because $m_j$ is constant (and is equal to the total number of grooms in that \jokbo{}).  The strongest correlation between the gravity-model flux and the number of entries for each clan in \jokbo{} 1 occurs for $\alpha \approx 1.0749$ and $\gamma \approx -0.0349$, which suggests that the frequency of marriage between two families is proportional to the product of the populations of the two clans and, in particular, that there is little or no geographical dependence. The likely explanation is that the clan in \jokbo{} 1 is ergodic, so the grooms could have been almost anywhere in the country, which would indeed make geographical factors irrelevant. (In the context of population genetics, this corresponds to ``full mixing''~\cite{Nei1966,Scapoli2007,Rossi2013,Ralph2013}.) 
In other words, as we discussed in Sec.~\ref{sec:generative_models}, this special case of the gravity model (for which we use $\gamma = 0$ in our analysis) corresponds to having geographical independence.  Consequently, we will henceforth use the term ``population-product model'' for the gravity model with $\gamma = 0$. For our analysis of other \jokbo{} and additional details, see Appendix~\ref{A:jokbo_data} (and Tables~\ref{table_num_ent} and \ref{jokbo1_temporally_divided}).

\begin{figure*}[t]
\includegraphics[width=0.7\textwidth]{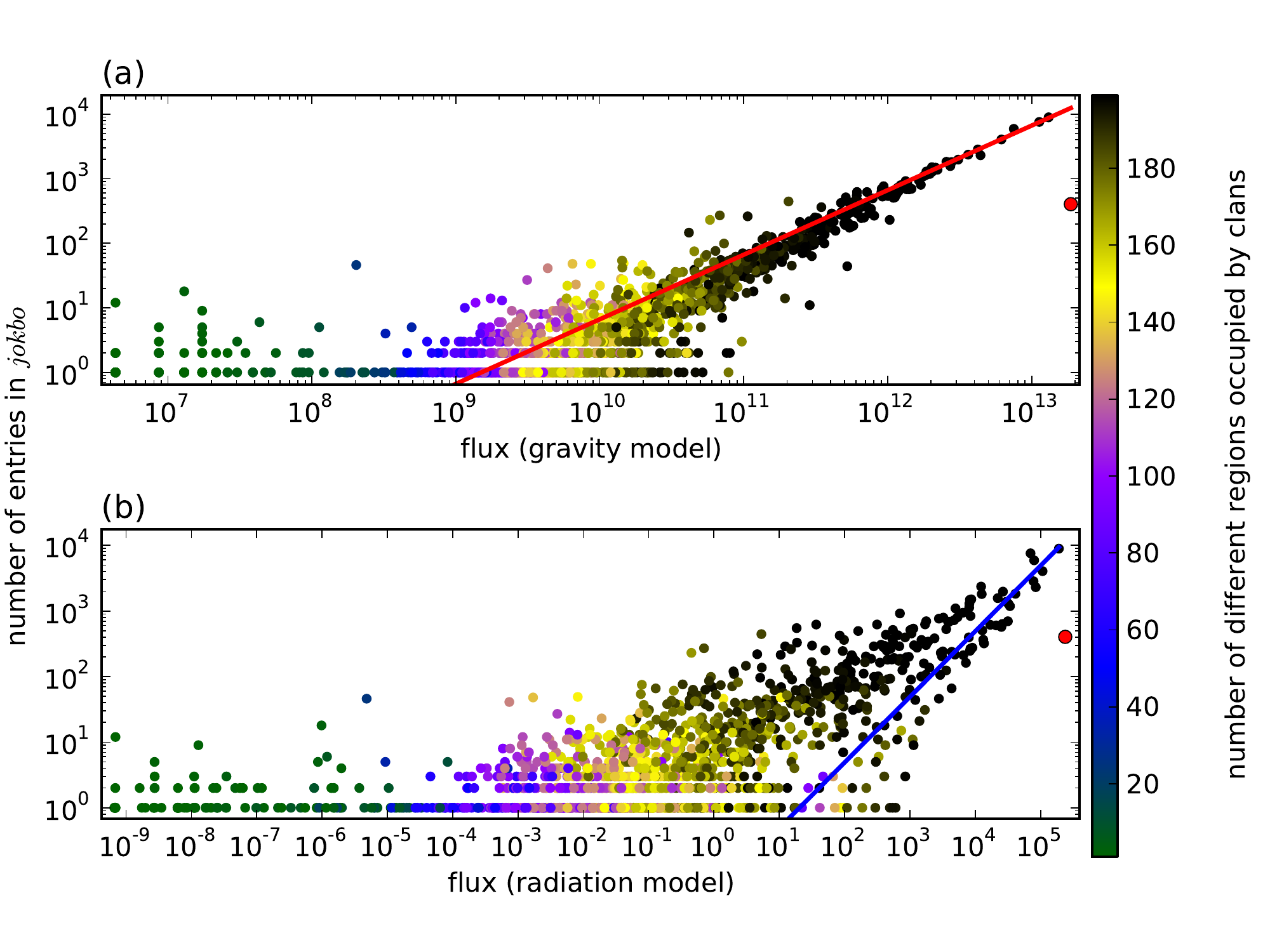}
\caption{
Flux predictions from the population-product model (i.e., the special case of the gravity model with $\gamma = 0$) with $\alpha = 1$ and the radiation models for \jokbo{} 1.
(a) Scatter plot of the number of clan entries in \jokbo{} 1 versus the corresponding centroid in 2000 using the population-product-model flux of women from clan $i$ to clan $j$ and with $\alpha = 1$. We compute the line using a linear regression to find the fitting parameter $a_G \approx 6.55(4) \times 10^{-11}$ (with a 95\% confidence interval) to satisfy the expression $N_i = a_G G_{ij}$, where $G_{ij}$ is the population-product-model flux and $N_i$ is the total number of entries from clan $i$ in the \jokbo{}.  (b) We compare the same clan entries using the radiation model. We compute the line using a linear regression to find the fitting parameter $a_R \approx 0.049(2)$ to satisfy the expression $N_i = a_R R_{ij}$, where $R_{ij}$ is the radiation-model flux of women from clan $i$ to clan $j$ and $N_i$ is the total number of entries from clan $i$ in the \jokbo{}. In both panels, we color the points using the number of administrative regions that are occupied by the corresponding clans [see Figs.~\ref{occu_num_diff_admin}(a) and (b)]. The red markers (outliers) in both panels correspond to the clan of \jokbo{} 1 (i.e., the case $i = j$).
}
\label{gravity_radiation_models_RG}
\end{figure*}

With little loss of accuracy for the fit, we take $\gamma = 0$ (i.e., we use the population-product model) to avoid divergence in the rare cases in which a bride comes from the same clan as the groom (for which the distance is $r_{ij}=0$).  We also take $\alpha = 1$ with little loss of accuracy.  Using $\gamma = 0$ allows us to include data from the approximately 22\% of clans for which geographical origin information is not available.  In Fig.~\ref{gravity_radiation_models_RG}, we show the fit for \jokbo{} 1, where we have used linear regression to quantify the correlation between the population-product-model flux and the number of entries for each clan in the \jokbo{}. The noticeably lower outlier to the right of the line is the data point that corresponds to the clan of \jokbo{} 1, and we remark that this deviation results from a cultural taboo against marrying into one's own clan. Women from the same clan as the owners of a \jokbo{} have traditionally been strongly discouraged from marrying men listed in the \jokbo{} (it is possible that they were even recorded under false clans in the book), and it was illegal until 1997~\cite{same_fam_marri_law}. For the other \jokbo{}, see Fig.~\ref{multi_jokbo_simplest_grav} in Appendix~\ref{A:jokbo_data}. In the bottom panel of Fig.~\ref{gravity_radiation_models_RG}, we illustrate that the radiation model does not give a good fit to the data. Recall from our discussion in Sec.~\ref{sec:generative_models} that the lack of parameters in the radiation model does not allow us to explicitly consider a geographically independent special case when using it. We emphasize, however, that this does not imply that the gravity model is ``better'' than the radiation model, as a direct comparison between the two models is hampered by the ergodicity of clans. In other words, the standard formulations of the gravity and radiation models do not provide a solution for how to estimate fluxes between the clan centroids. Consequently, to investigate population fluxes, we incorporate modern census data.  See our discussions in the next subsection and in Appendix \ref{sec:centrality_of_seoul}.

\subsection{Ergodicity analysis based on modern census data and a simple diffusion model}

We use census data from the year 2000~\cite{population_census} to examine the ergodicity of clans in three different ways: (1) The number of administrative regions quantifies how ``widely'' each clan is distributed; (2) the radius of gyration, which we calculate from the clan-density anomaly using Eq.~\eqref{radius_of_gyration}, quantifies how ``uniformly'' each clan is distributed; and (3) the standard deviation of anomaly value measures how much the anomaly varies across regions. For instance, using data from the 2000 census and considering all of the clans and the $199$ standardized regions, we find that $3.04\%$ of the clans have a member in every region but that $22.1\%$ of the clans have members in ten or fewer regions.

\begin{figure}[t]
\includegraphics[width=0.9\columnwidth]{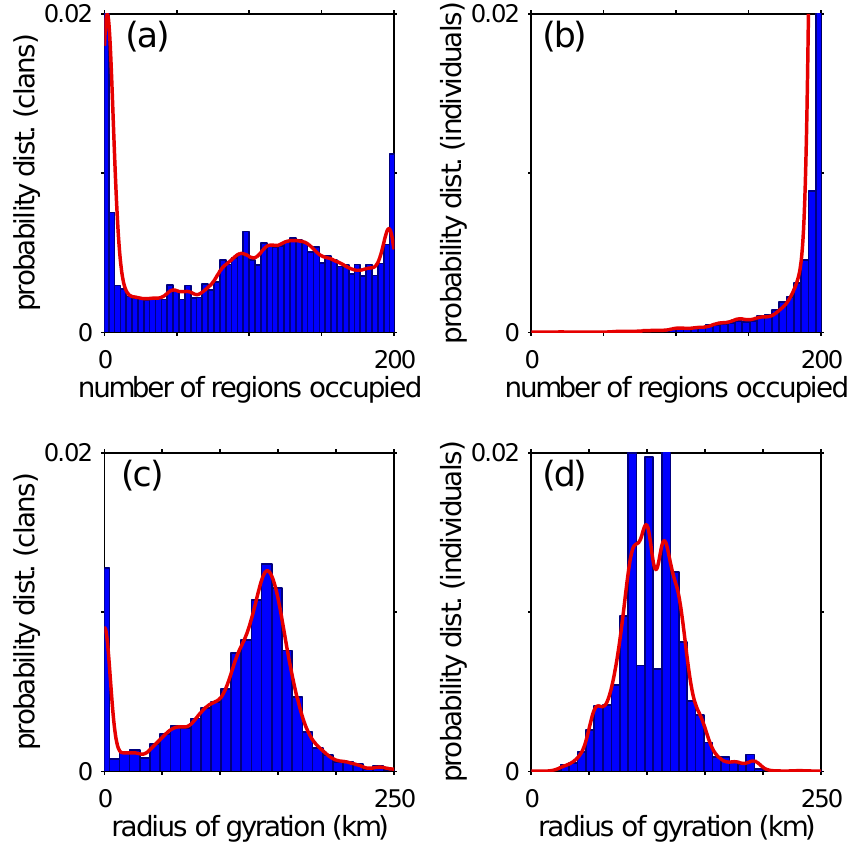} 
\caption{Distribution of the number of different administrative regions occupied by clans. (a) Probability distribution of the number of different administrative regions occupied by a Korean clan in the year 2000. (b) Probability distribution of the number of different administrative regions occupied by the clan of a Korean individual selected uniformly at random in the year 2000.  The difference between this panel and the previous one arises from the fact that clans with larger populations tend to occupy more administrative regions. [That is, we select a clan uniformly at random in panel (a), but we select an individual uniformly at random in panel (b).] Note that the rightmost bar has a height of $0.17$, but we has truncated it for visual presentation. (c) Probability distribution of radii of gyration (in km) for clans in 2000. (d) Probability distribution of radii of gyration (in km) for clans of a Korean individual selected uniformly at random in 2000. The difference between this panel and the previous one arises from the fact that clans with larger populations tend to occupy more administrative regions. 
Solid curves are kernel density estimates (from {\sc Matlab} R2011a's \texttt{ksdensity} function with a Gaussian smoothing kernel of width 5).
}
\label{occu_num_diff_admin}
\end{figure}

We illustrate the dichotomy of ergodic versus nonergodic clans with the bimodal distribution in Fig.~\ref{occu_num_diff_admin}(a).  However, from the perspective of individual clan members [see Fig.~\ref{occu_num_diff_admin}(b)], such a dichotomy is not apparent.  We show the radii of gyration that we calculate from the 2000 census data in Figs.~\ref{occu_num_diff_admin}(c) and ~\ref{occu_num_diff_admin}(d).  We can again see the bimodality in Fig.~\ref{occu_num_diff_admin}(c). In Fig.~\ref{phi_distribution_for_Kim_and_Lee} in Appendix~\ref{sec:other_results}, we illustrate the dichotomy for Kim from Gimhae and Lee from Hakseong.

As we indicate in Table~\ref{table_num_ent}, all ten of the clans for which we have \jokbo{} are ergodic or at least reasonably ergodic, so the variables associated with the $j$ indices (i.e., the grooms) in Eqs.~\eqref{grav_mod_eq} and \eqref{rad_mod_eq} have already lost much of their geographical precision, which is consistent with both $\gamma = 0$ (i.e., with using the population-product model) and $\alpha = 0$.   
Again see the scatter plots in Fig.~\ref{gravity_radiation_models_RG}, in which we color each clan according to the number of different administrative regions that it occupies. Note that the three different ergodicity diagnostics are only weakly correlated with each other (see Fig.~\ref{distance_num_admin_RG} in Appendix~\ref{sec:other_results}).

Our observations of clan bimodality for Korea contrast sharply with our observations for family names in the Czech Republic, where most family names appear to be nonergodic~\cite{czech_surname_pap} (see Fig.~\ref{occu_num_diff_admin_Czech} in Appendix~\ref{sec:other_results}). One possible explanation of the ubiquity of ergodic Korean names is the historical fact that many families from the lower social classes adopted (or even purchased) names of noble clans from the upper classes near the end of the Joseon dynasty (19th--20th centuries){~\cite{KoreanFamilyNamesAndGenealogies,JoseonDynasty}}. At the time, Korean society was very unstable, and this process might have, in essence, introduced a preferential growth of ergodic names.

\begin{figure}[t]
\includegraphics[width=0.9\columnwidth]{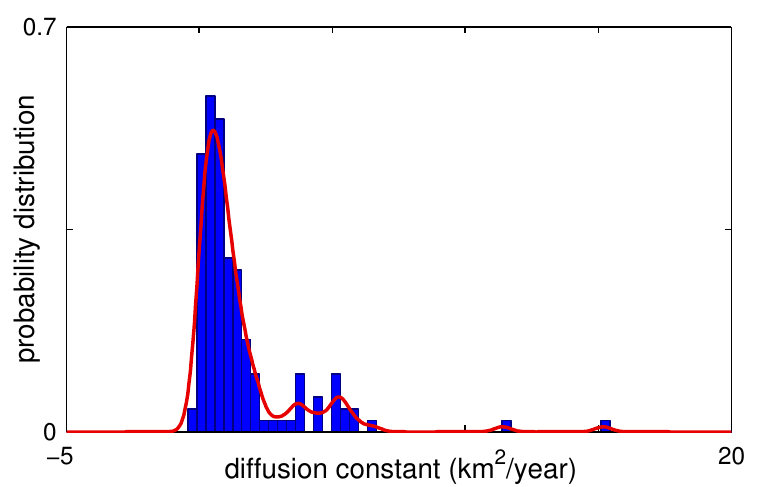}
\caption{Distribution of estimated diffusion constants (in km$^2$/year) computed using 1985 and 2000 census data and Eq.~\eqref{diffusion_eq}. The solid curve is a kernel density estimate (from {\sc Matlab} R2011a's \texttt{ksdensity} function with default smoothing). See the Appendix~\ref{sec:estimating_diffusion_constants} for details of the calculation of diffusion constants.
}
\label{diffusion_constant_distribution}
\end{figure}

In Fig.~\ref{diffusion_constant_distribution}, we show the distribution of the diffusion constants that we computed by fitting to Eq.~\eqref{diffusion_eq}.  Some of the values are negative, which presumably arises from finite-size effects in ergodic clans as well as basic limitations in estimating diffusion constants using only a pair of nearby years.  In Fig.~\ref{diffusion_distance_num_admin_RG} in Appendix~\ref{sec:other_results}, we show the correlations between the diffusion constants and other measures.

\subsection{Convection in addition to diffusion as another mechanism for migration}
The assumption that human populations simply diffuse is a gross oversimplification of reality.  We will thus consider the intriguing (but still grossly oversimplified) possibility of simultaneous diffusive and convective (bulk) transport. In the past century, a dramatic movement from rural to urban areas has caused Seoul's population to increase by a factor of more than 50, tremendously outpacing Korea's population growth as a whole~\cite{SeoulPopulationHistory}.  This suggests the presence of a strong attractor or ``sink'' for the bulk flow of population into Seoul, as has been discussed in rural-urban labor migration studies~\cite{Todaro1969}. The density-equalizing population cartogram~\cite{Gastner2004} in Fig.~\ref{cartogram} in Appendix~\ref{sec:other_results} clearly demonstrates the rapid growth of Seoul and its surroundings between 1970 and 2010.

\begin{figure}[t]
\includegraphics[width=0.9\columnwidth]{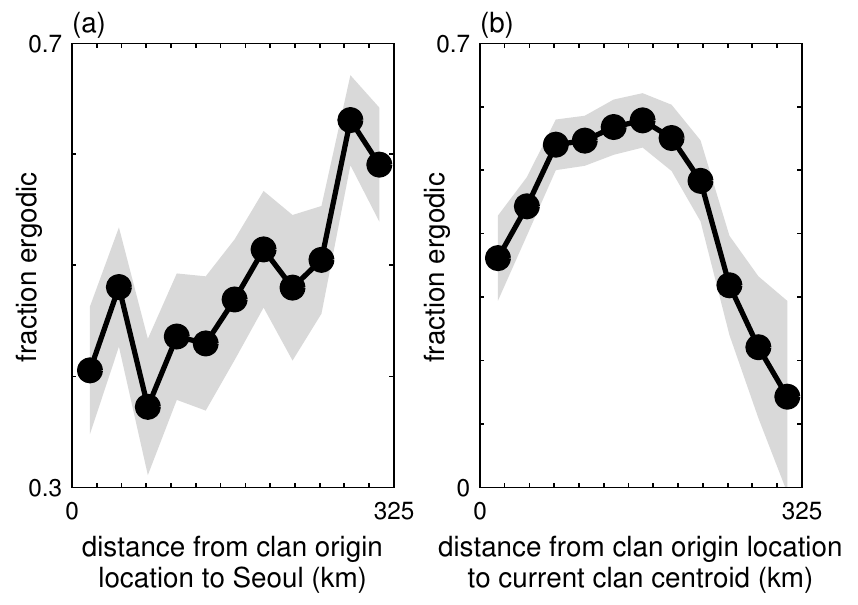} 
\caption{
Fraction of ergodic clans and distance scales of clans. For this figure, we use the 3900 clans from the 2000 census data for which we were able to identify the origin locations (see Appendix~\ref{sec:geographical_information_from_google_maps}).
 (a) Fraction of ergodic clans versus distance to Seoul.  The correlation between the variables is positive and statistically significant. (The Pearson correlation coefficient is $r \approx 0.83$, and the $p$-value is $p \approx 0.0017$.)  For the purpose of this calculation, we call a clan ``ergodic'' if it is present in at least 150 administrative regions. We estimate this fraction separately in each of 11 equally sized bins for the displayed range of distances.  The gray regions give 95\% confidence intervals.  (b) Fraction of ergodic clans versus the distance between the location of clan origin and the present-day centroid.  We measure ergodicity as in the left panel, and we estimate the fraction separately for each range of binned distances. (We use the same bins as in the left panel.) The correlation between the variables is positive and significant up to 150 km ($r \approx 0.94$, $p \approx 0.0098$) and is negative and significant for larger distances ($r \approx -0.98$, $p \approx 2.4 \times 10^{-4}$). 
}
\label{erg_vs_dist_fig}
\end{figure}

If convection (i.e., bulk flow) directed towards Seoul has indeed occurred throughout Korea while clans were simultaneously diffusing from their points of origin, then one ought to be able to detect a signature of such a flow.  In Fig.~\ref{erg_vs_dist_fig}(a), we show what we believe is such a signature. We observe that the fraction of ergodic clans increases with the distance between Seoul and a clan's place of origin.  This would be unexpected for a purely diffusive system or, indeed, in any other simple model that excludes convective transport.  By allowing for bulk flow, we expect to observe that a clan's members preferentially occupy territory in the flow path that is located geographically between the clan's starting point and Seoul.  For clans that start closer to Seoul, this path is short; for those that start farther away, the longer flow path ought to contribute to an increased number of occupied administrative regions and hence to a greater aggregate ergodicity. We plot the fraction of ergodic clans versus the distance a clan has moved (which we estimate by calculating distances between clan origin locations and the corresponding modern clan centroids) in Fig.~\ref{erg_vs_dist_fig}(b). This also supports our claim that both convective and diffusive transport have occurred.  To further examine clan ergodicity, we also compare each clan's radius of gyration $r_g$ to the distance from its origin location to (1) Seoul and (2) its present-day centroid (see Fig.~\ref{r_g_vs_dist_fig} in Appendix~\ref{sec:other_results}).  The latter shows the same general tendency as in Fig.~\ref{erg_vs_dist_fig}.
 We speculate that the absence of statistical significance in the correlation between $r_g$ and the distances between clan origin locations and Seoul is a sampling issue, as we could not include many of the small clans in this calculation because we cannot estimate the locations of their centroids from our data (see Appendix~\ref{sec:census}).

We assume that clans that have moved a larger distance have also existed for a longer time and hence have undergone diffusion longer; we thus also expect such clans to be more ergodic.  This is consistent with our observations in Fig.~\ref{erg_vs_dist_fig}(b) for distances less than about 150 km, but it is difficult to use the same logic to explain our observations for distances greater than 150 km. However, if one assumes that long-distance moves are more likely to arise from convective effects than from diffusive ones, then our observations for both short and long distances become understandable. The fraction of moves from bulk-flow effects like resettlement or transplantation is larger for long-distance moves, and they become increasingly dominant as the distance approaches 325 km (roughly the size of the Korean peninsula).  We speculate that the clans that moved farther than 150 km are likely to be ones that originated in the most remote areas of Korea, or even outside of Korea, and that they have only relatively recently been transplanted to major Korean population centers, from which they have had little time to spread.  This observation is necessarily speculative because the age of a clan is not easy to determine. The first entry in a \jokbo{} (see Table~\ref{table_num_ent} for our ten \jokbo{}) could have resulted from the invention of characters or printing devices rather than from the true birth of a clan{~\cite{KoreanFamilyNamesAndGenealogies}}.

Ultimately, our data are insufficient to definitively accept or reject the hypothesis of human diffusion.  However, as our analysis demonstrates, our data are consistent with the theory of simultaneous human ``diffusion'' and ``convection''.  Furthermore, our analysis suggests that \textit{if} the hypothesis of pure diffusion is correct, then our estimated diffusion constants indicate a possible time scale for relaxation to a dynamic equilibrium and thus for mixing in human societies. In mainland South Korea, it would take approximately $(100\,000 \textrm{ km}^2)/(1.5 \textrm{ km}^2/\textrm{year}) \approx 67\,000$ years for purely diffusive mixing to produce a well-mixed society.  A convective process thus appears to be playing the important role of promoting human interaction by accelerating  mixing in the population.  Despite the limitations imposed by our data, we try to estimate and quantify the centrality of Seoul using a network-flow model for population, and we find suggestive differences between the flow patterns of ergodic and nonergodic clans. For details, see Appendix~\ref{sec:centrality_of_seoul}.

\section{Conclusions}
\label{sec:conclusions}

The long history of detailed record-keeping in Korean culture provides an unusual opportunity for quantitative research on historical human mobility and migration, and our investigation strongly suggests that both ``diffusive'' and ``convective'' patterns have played important roles in establishing the current distribution of clans in Korea.  By studying the geographical locations of clan origins in \jokbo{} (Korean family books), we have quantified the extent of ``ergodicity'' of Korean clans as reflected in time series of marriage snapshots. This underscores the utility of investigating the location distributions of individual clans.
Additionally, by comparing our results from Korean clans to those from Czech families, we have also demonstrated that our approach can give insightful 
indications of different mobility and migration patterns in different cultures.  Our ergodicity analysis using modern census data clearly illustrates that there are both ergodic and nonergodic clans, and we have used these results to suggest two different mechanisms for human migration on long time scales.
Many mobility processes involve a balance between diffusive spreading and attraction to one or more central locations (and between more general diffusive and convective fluxes), so we believe that our approach in the present paper will be valuable for many situations. 

A noteworthy feature of our analysis is that we used both data with high temporal resolution but low spatial resolution (\jokbo{} data) and data with high spatial resolution but low temporal resolution (census data).  This allowed us to consider both the patterns of human movement on short time scales (mobility via individual marriage processes) and their consequences for human locations on long time scales (human migration via clan ergodicity).  An interesting further wrinkle would be to compare such mobility-derived time scales for human mixing patterns to genetically-derived patterns~\cite{Nei1966,Scapoli2007,Rossi2013,Ralph2013}.

From a more general perspective, our research has allowed us to test the idea of using a physical analogy for modeling human \emph{migration}---an idea put forth (but not quantified) as early as the 19th century \cite{Ravenstein1885,Stouffer1940,orig_grav_mobil_pap,Wilson1967,Thornthwaite1934,Lucas1981,Dorigo1983,Greenwood2003,Stewart1950,Quetelet1835}.
Physics-inspired ideas have been very successful for the study of human \emph{mobility}, which occurs on shorter time scales than human migration, and we propose that Ravenstein was correct when he posited that such ideas are also useful for human migration.

\begin{acknowledgments}
We thank Hawoong Jeong (정하웅) for providing data from Korean family books and Josef Novotn\'{y} for providing data on surnames in the Czech Republic.  We thank Tim Evans for introducing us to helpful references and Erik Bollt, Valentin Danchev, Sandra Gonz\'{a}lez-Bail\'{o}n, James Irish, Philip Kreager, Michael Murphy, and Tommy Murphy for helpful comments and discussions. We thank Marc Barthelemy and Richard Morris for details about their work on constructing flow networks~\cite{Morris2012}, and we thank the anonymous referees for their helpful comments and suggestions. M.\,A.\,P. and S.\,H.\,L. acknowledge support from the Engineering and Physical Sciences Research Council (EPSRC) through grant No. EP/J001759/1. B.\,J.\,K. was supported by a National Research Foundation of Korea (NRF) grant funded by the Korean government (No. 2014R1A2A2A01004919). D.\,M.\,A. was supported by Grant No. \#220020230 from the James S. McDonnell Foundation. S.\,H.\,L. did the majority of his work at University of Oxford.
\end{acknowledgments}

\appendix

\section{Jokbo Data}
\label{A:jokbo_data}

\begin{figure*}[t]
\includegraphics[width=0.8\textwidth]{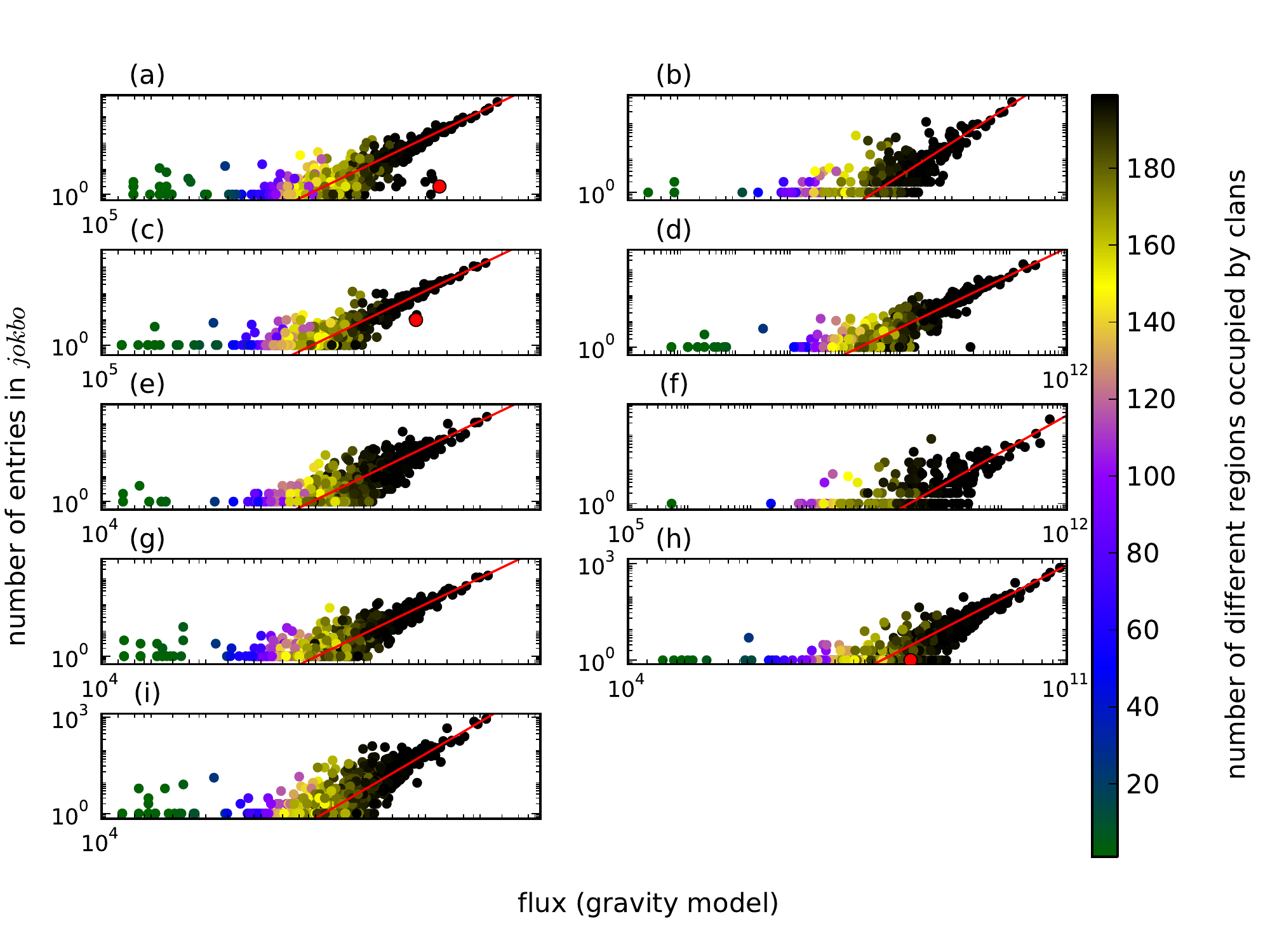}
\caption{Scatter plots of the number of clan entries in \jokbo{} 2--10 
versus the corresponding centroid in 2000 using the population-product-model flux with 
$\alpha = 1$.
 We show our results in numerical order of the \jokbo{} in panels (a)--(i), so \jokbo{} 2 is in panel (a), etc.  In each panel, we calculate the line using linear regression to determine the fitting parameter $a_G$ for $N_i = a_G G_{ij}$, where $G_{ij}$ is the population-product-model flux of women from clan $i$ to clan $j$ and $N_i$ is the total number of entries from clan $i$ in the given \jokbo{}.  The parameter values are
(a) $a_G \approx 2.36(1) \times 10^{-9}$ [\jokbo{} 2], 
(b) $a_G \approx 6.6(1) \times 10^{-11}$ [\jokbo{} 3], 
(c) $a_G \approx 5.15(5) \times 10^{-9}$ [\jokbo{} 4],
(d) $a_G \approx 5.15(5) \times 10^{-8}$ [\jokbo{} 5],
(e) $a_G \approx 5.8(1) \times 10^{-9}$ [\jokbo{} 6],
(f) $a_G \approx 5.1(2) \times 10^{-16}$ [\jokbo{} 7], 
(g) $a_G \approx 4.25(5) \times 10^{-8}$ [\jokbo{} 8], 
(h) $a_G \approx 1.44(1) \times 10^{-9}$ [\jokbo{} 9], and 
(i) $a_G \approx 4.71(8) \times 10^{-8}$ [\jokbo{} 10].
The red markers in panels (a), (c), and (h) correspond to the clans of the depicted \jokbo{}, and $N_i |_{i = j = \textrm{own clan}} = 0$ for all of the other \jokbo{}.
In each case, we use a 95\% confidence interval and color the points according to the
number of administrative regions occupied by the corresponding clans.}
\label{multi_jokbo_simplest_grav}
\end{figure*}

\begin{figure*}[t]
\includegraphics[width=0.8\textwidth]{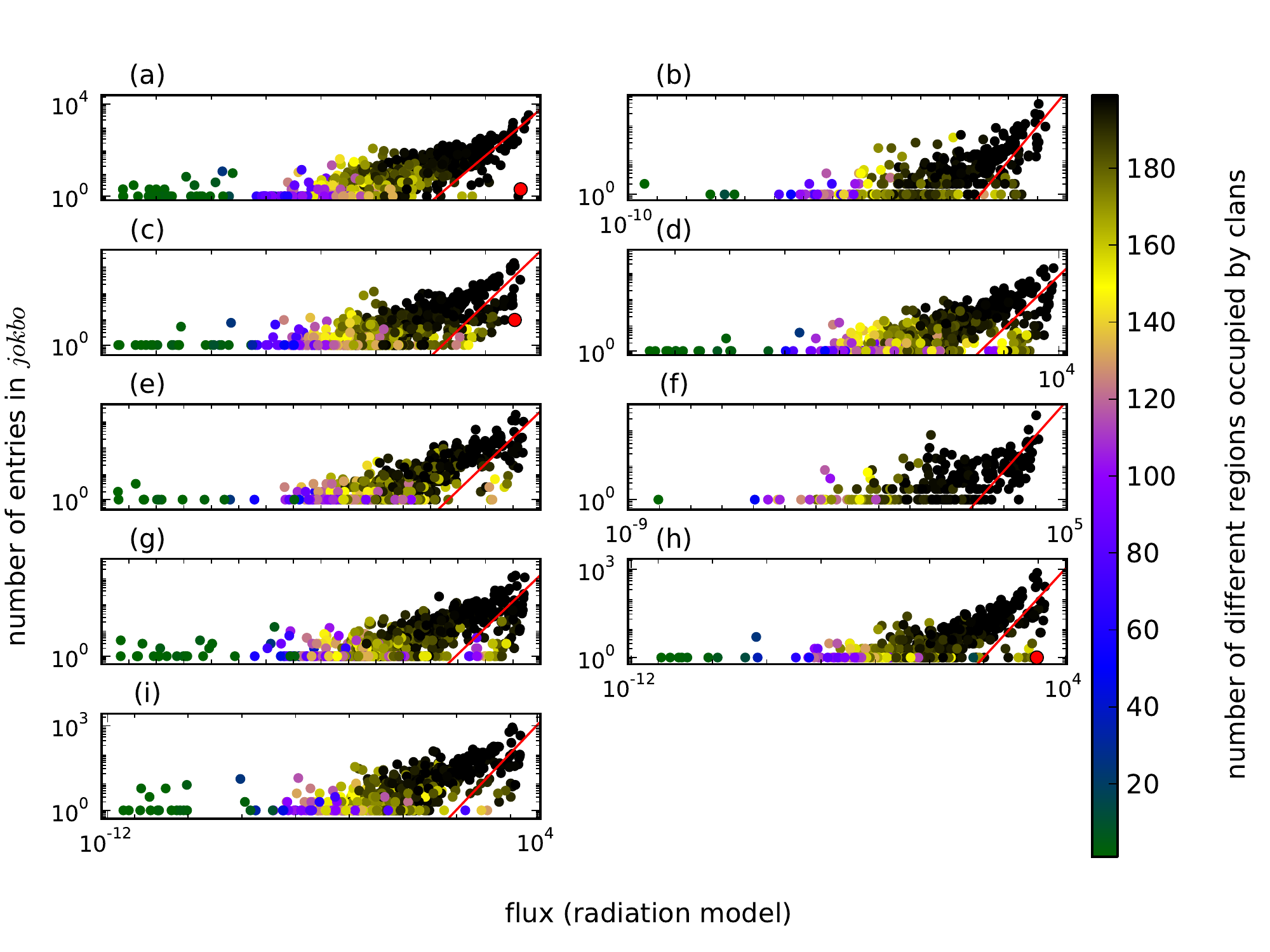}
\caption{Scatter plots of the number of clan entries in \jokbo{} 2--10 versus the corresponding centroid in 2000 using the radiation-model flux.  We show our results in numerical order of the \jokbo{} in panels (a)--(i), so \jokbo{} 2 is in panel (a), etc. In each panel, we calculate the line using a linear regression to determine the fitting parameter $a_R$ for $N_i = a_R R_{ij}$, where $R_{ij}$ is the radiation-model flux of women from clan $i$ to clan $j$ and $N_i$ is the total number of entries from clan $i$ in the \jokbo{}. The parameter values are
(a) $a_R \approx 0.062(2)$ [\jokbo{} 2], 
(b) $a_R \approx 0.0098(7)$ [\jokbo{} 3], 
(c) $a_R \approx 0.040(3)$ [\jokbo{} 4],
(d) $a_R \approx 0.075(7)$ [\jokbo{} 5],
(e) $a_R \approx 0.23(2)$ [\jokbo{} 6],
(f) $a_R \approx 0.0069(5)$ [\jokbo{} 7], 
(g) $a_R \approx 0.12(1)$ [\jokbo{} 8], 
(h) $a_R \approx 0.11(1)$ [\jokbo{} 9], and 
(i) $a_R \approx 0.11(1)$ [\jokbo{} 10].
The red markers in panels (a), (c), and (h) correspond to the clans of the depicted \jokbo{}, and $N_i |_{i = j = \textrm{own clan}} = 0$ for all of the other \jokbo{}.
In each case, we use a 95\% confidence interval and color the points according to the number of administrative regions occupied by the corresponding clans.}
\label{multi_jokbo_rad}
\end{figure*}

In our investigation, we examine ten digitized \jokbo{} that were first studied in Ref.~\cite{Kiet2007}. In Table \ref{table_num_ent} in the main text, we give basic information about the ten \jokbo{} and we now summarize the results of some of our computations. 

First, we apply the same gravity-model fit that we used for \jokbo{} 1 to all of the \jokbo{} data, and the results do not deviate much from those for \jokbo{} 1.  That is, $\gamma \approx 0$ and $\alpha \approx 1$, so we can apply the population-product model with $\alpha = 1$.  The largest deviations in the two parameter values are $\alpha \approx 1.4930$ (for \jokbo{} 7) and $\gamma \approx 0.5377$ (for \jokbo{} 6). Interestingly, we could not find \emph{any} empirical value of $\gamma < 0.6$ reported in the literature \cite{orig_grav_mobil_pap, Wilson1967, gravvsradmodel, Young1924, highw_grav_pap, subwa_grav_pap, diseases_grav_pap}, and it seems to be extremely rare to report any empirical values at all for gravity-model parameters. As one can see in Fig.~\ref{multi_jokbo_simplest_grav}, the choice of $\alpha = 1$ and $\gamma = 0$ fits the data reasonably well for \jokbo{} 2--10. Note that the suppressed case of a bride and groom being from the same clan is apparent in Fig.~\ref{multi_jokbo_simplest_grav}.  This is indicated by the red markers, which are significantly below the other points in some of the panels and do not exist at all in other panels. We show the radiation-model results for other \jokbo{} 2--10 in Fig.~\ref{multi_jokbo_rad}. 

\begin{table}[t]
\caption{Gravity-model parameters $\alpha$ and $\gamma$ in Eq.~\eqref{grav_mod_eq} calculated using the clan origin locations (instead of the population centroids from the census data) for all of the \jokbo{} by minimizing the sum of squared differences using the \texttt{scipy.optimize} package in {\sc Python}~\cite{ASciPyOptimization}. (We again use initial values of $\alpha = \gamma = a_G = 1.0$ for \jokbo{} 1--2, 4--6, and 8--10.  For \jokbo{} 3 and 7, we instead use $\alpha = a_G = 1.0$ and $\gamma = 0.01$ because the procedure did not converge when we used $\alpha = \gamma = a_G = 1.0$.)}
\begin{tabular}{rrr}
\hline
\jokbo{} ID & $\alpha$ & $\gamma$ \\
\hline
1 & $1.1188$ & $-0.0716$ \\
2 & $1.1261$ & $0.1253$ \\
3 & $1.0983$ & $0.0205$ \\
4 & $0.9310$ & $-0.1441$ \\
5 & $0.8922$ & $0.0540$ \\
6 & $0.9785$ & $0.2776$ \\
7 & $1.5820$ & $0.0183$ \\
8 & $0.9651$ & $0.1285$ \\
9 & $1.1683$ & $0.3600$ \\
10 & $0.9244$ & $0.0868$ \\
\hline
\end{tabular}
\label{gravity_model_parameter_with_origin}
\end{table}

\begin{figure*}[t]
\includegraphics[width=0.7\textwidth]{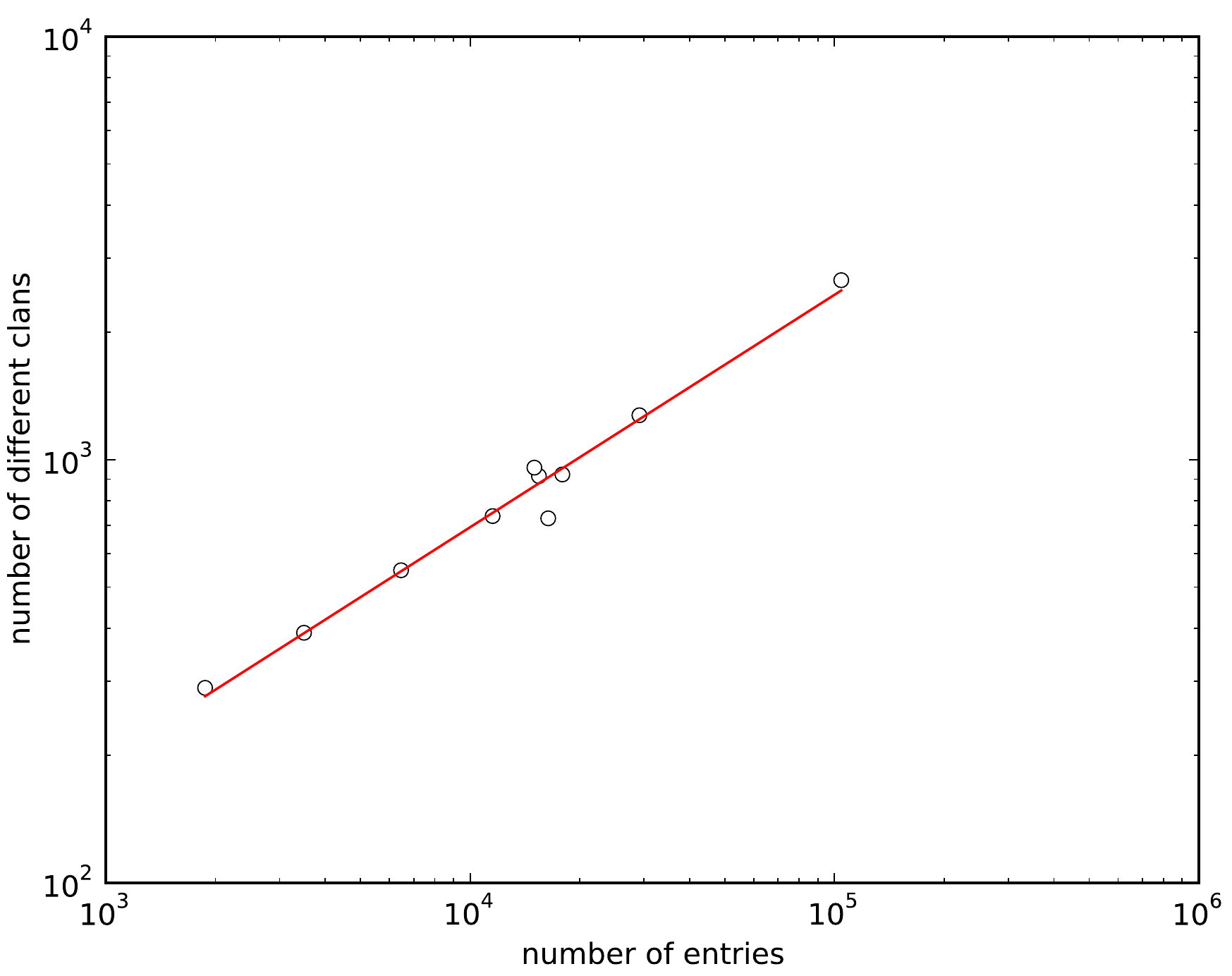}
\caption{For each \jokbo{}, we plot the number of distinct clans $N_c$ versus the total number of entries $N_e$ on a doubly logarithmic scale.  
We calculate the red line via a linear regression to Heaps' law~\cite{AHeapsLaw} 
using the expression $N_c = 10^{b_J} N_e^{a_J}$.  This yields a slope of $a_J \approx 0.55(7)$ and an intercept of $b_J \approx 0.6(3)$ (with 95\% confidence intervals).
}
\label{num_ent_num_fam_fig}
\end{figure*}

\begin{figure*}[t]
\begin{tabular}{ll}
{\sffamily (a)} & {\sffamily (b)} \\ 
\includegraphics[width=0.48\textwidth]{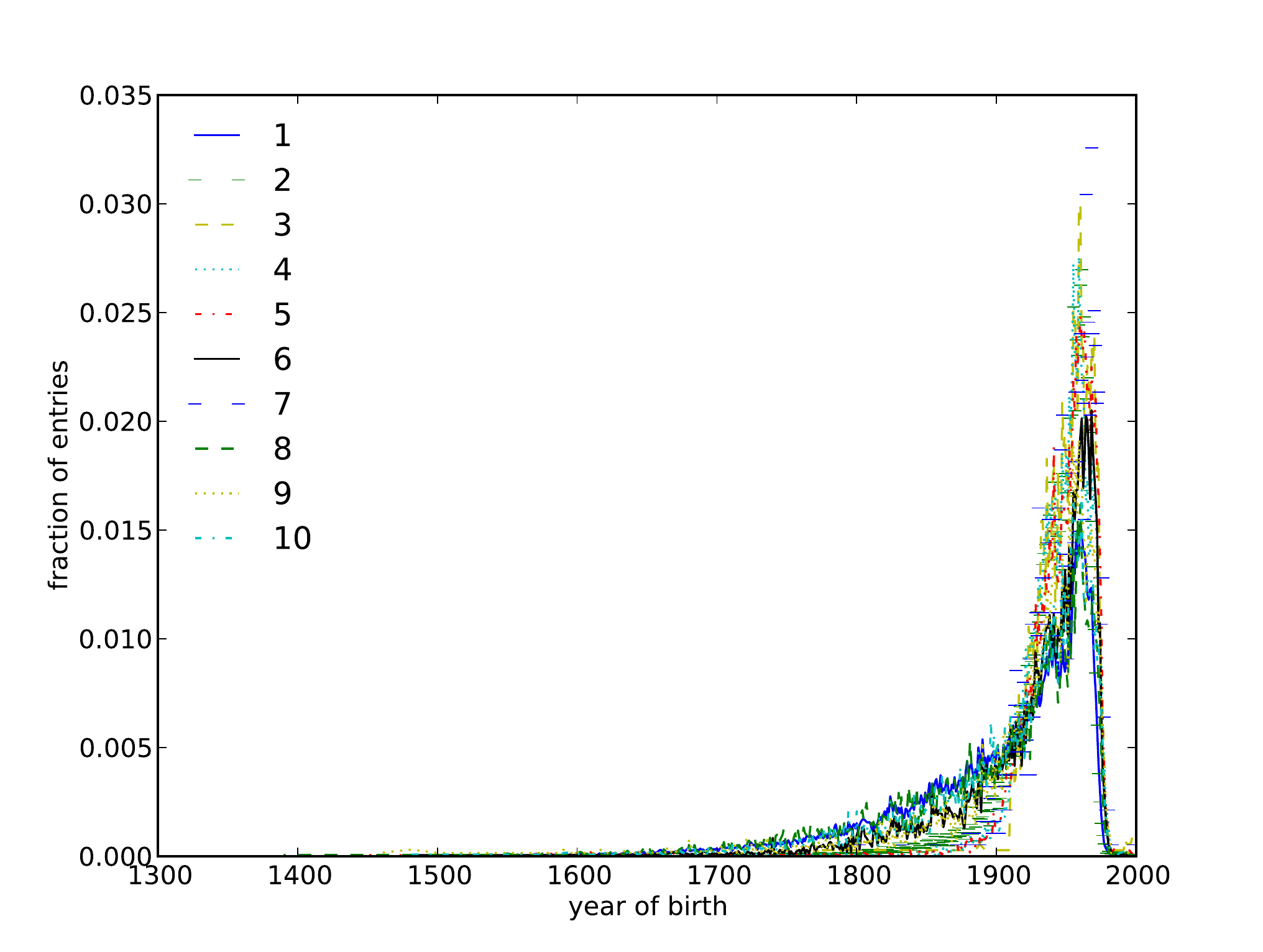} &
\includegraphics[width=0.48\textwidth]{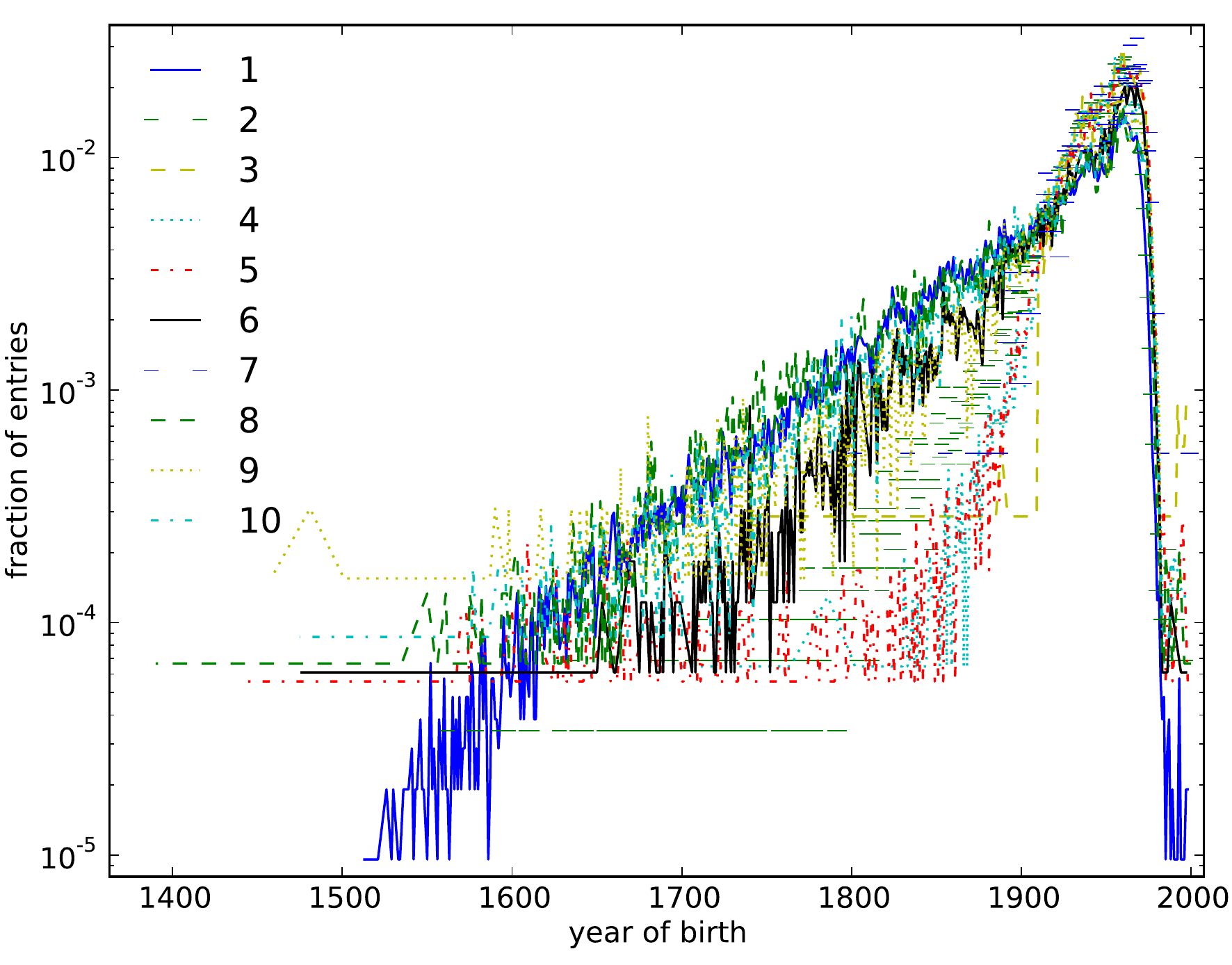} \\
\end{tabular}
\caption{Fraction of entries in each \jokbo{} as a function of the birth year of brides using (a) linear and (b) semilogarithmic scales.
The sudden drop on the right of each panel simply reflects the fact that women who are too young are not yet married.
}
\label{collaps_num_ent_year_linear}
\end{figure*}

Additionally, we can see that all of the clans in the \jokbo{} data that we study (i.e., the grooms' side of marriages) are ``ergodic'' in the sense that they were widespread across the nation in 2000. This is not surprising, as the availability of digitized \jokbo{} data themselves reflects clan popularity. We present the gravity-model fitting results for temporally-divided \jokbo{} 1 data in Table~\ref{jokbo1_temporally_divided} in the main text, and we give results that use clan origin locations instead of population centroid in 2000 in Table~\ref{gravity_model_parameter_with_origin}.  (We also temporally-divided the data from \jokbo{} 6---because, as we showed in Table~\ref{table_num_ent} in the main text, it has the largest $\gamma$ value among the ten clans---and we found that it does not exhibit systematic changes over time either.)
With these calculations, we again find that $\alpha \approx 1$ and $\gamma \approx 0$ appear to be reasonable. The general trend of population change in Korea is also reflected in the \jokbo{} data.
In Fig.~\ref{num_ent_num_fam_fig}, we examine the number of distinct clans in each \jokbo{} versus the total number of entries in that \jokbo{}. In Fig.~\ref{collaps_num_ent_year_linear}, we show the fraction of entries in each \emph{jokbo} as a function of the birth year of the brides in that \jokbo{}. These plots suggest that \jokbo{} of different sizes at different times tend to follow the aggregate trend of population change throughout the last several hundred years of Korean history.

\section{Census Data, Populations, and Numbers of Clans}  
\label{sec:census}

Since 1925, the South Korean government has conducted a census every five years \cite{population_census}. The only years in which the populations of different clans were recorded separately for different administrative regions were 1985 and 2000.  These data make it possible to estimate distribution statistics (e.g., centroid and radius of gyration) for each clan. All of the data are publicly available to download from Ref.~\cite{population_census}.

The total population reported in the 1985 South Korean census was 40\,419\,647, and clan information is available for 40\,315\,813 individuals.  In the 2000 South Korean census,  a population of 45\,985\,289 was reported, and a clan is indicated for every individual. The number of different clans identified in the 1985 (respectively, 2000) census was 3\,520 (respectively, 4\,303). There are 3\,481 clans in common in the two censuses: 39 clans disappeared and 822 new ones appeared.  

\begin{figure}[t]
\includegraphics[width=0.9\columnwidth]{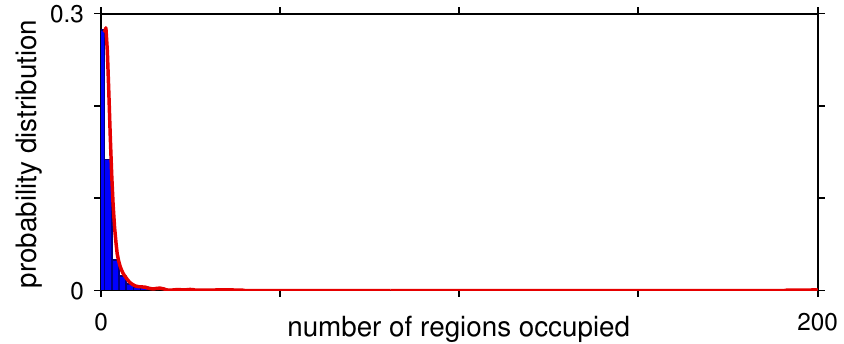}
\caption{Probability distribution for the number of different administrative regions occupied by the 822 ``new'' clans that are in the 2000 census data but are not in the 1985 census data. The solid curve is a kernel density estimate (from {\sc Matlab} R2011a's \texttt{ksdensity} function with a Gaussian smoothing kernel of width 1).} 
\label{occu_num_diff_admin_new_2000}
\end{figure}

In Fig.~\ref{occu_num_diff_admin_new_2000}, we indicate how many administrative regions the 822 ``new'' clans occupy. New clans might correspond to foreigners who obtained South Korean citizenship during the 15-year period 1985--2000, or these clans might simply have been missing erroneously from the 1985 census. Figure \ref{occu_num_diff_admin_new_2000} supports the hypothesis that these are genuinely new clans because their members have not spread to a large number of administrative regions.  This gives a total of 6\,687 distinct clans after we also incorporate the 2\,384 additional clans that are listed only in the \jokbo{}. In Table \ref{table_num_ent} in the main text, we indicate the number of distinct clans in each of the ten \jokbo{}.  There are 162 clans that appear in all ten \jokbo{}.  For all calculations with the gravity and radiation models, we use the 4\,303 clans listed in the 2000 census data. When we use the population-product model (for which $\gamma = 0$), we do not require geometrical information, so we also use the additional clans listed in each \jokbo{}.
In this case, we denote the number of clans by $N_{\gamma=0}$ (see Table~\ref{table_num_ent}).

\section{Standardizing Administrative Regions in 1985 and 2000}
\label{sec:standardized_regions}

For the administrative regions, we use municipal divisions that are composed of city (시 in Korean), county (군 in Korean), and district (구 in Korean)~\cite{Amunicipal_level_divisions}.  In South Korea, there were 232 (respectively, 246) such administrative regions in 1985 (respectively, 2000).  The difference in the number of regions between the two years reflects a slight restructuring of the political units. 

For our computations, we need to unify the two different partitionings to be able to systematically compare results from 1985 and 2000 and to compute diffusion constants. To do this, we manually extract 199 ``standardized'' regions that we use for all computations involving administrative regions.  Our construction necessitates many instances of operations like the following:
\begin{itemize}
\item{$A + B$ (1985) $\to$ $C$ (2000) $\Rightarrow$ $C$ (standardized region)}
\item{$A$ (1985) $\to$ $B + C$ (2000) $\Rightarrow$ $A$ (standardized region)}
\item{$A + B$ (1985) $\to$ $C + D + E$ (2000) $\Rightarrow$ $F$ (renamed standardized region)}
\end{itemize}
For each operation, the region on the right is the standardized one that we use in our computations.  In a given example, each different region is represented by a different letter.  Thus, in example (i), two distinct regions from the 1985 census have merged into one region (and correspond exactly to that region) from the 2000 census, and we use this last region as one of our 199 standardized regions.  In other situations, such as in example (iii), the standardized region does not correspond exactly to a single region from either census.  Finally, we remark that the above operations are examples of what we needed to do to reconcile the 1985 and 2000 administrative regions.  This is not an exhaustive list (e.g., four regions in 1985 corresponding to six regions in 2000), and we treat these other cases similarly.

For each standardized region, we sum the associated areas and populations of the constituent regions to obtain the area and population values that we use in our computations.   
We have posted the data for the standardized regions as 
``\textsc{standardized\_regions.txt}'' in the Supplemental Material~\cite{SM}. 
For each standardized region, these data include the component region names (in Korean) in 1985 and 2000, the latitudes and longitudes [and Universal Transverse Mercator (UTM) easting and northing coordinates; see Appendix~\ref{sec:UTM}] of the component region administrative centers, the geographical areas of the component regions, and the populations of the component regions in 1985 and 2000. The data are in a tab-delimited text file, for which we have used the 16-bit Unicode Transformation Format (UTF-16) encoding scheme~\cite{AUTF-16} for the Korean characters.

The regional boundaries drawn in Fig.~\ref{GimhaeKim_vs_HakseongLee} are from the 2010 data downloaded from Ref.~\cite{regional_boundary}. There is a slight difference between the regional boundaries in 2000 and 2010, so we map the coordinates of administrative regions in 2000 to those in 2010 by checking which ``polygon'' in 2010 encloses the coordinates of administrative regions from 2000. 

\section{Obtaining Geographical Information from Google Maps}
\label{sec:geographical_information_from_google_maps}

\begin{figure}[t]
\includegraphics[width=0.9\columnwidth]{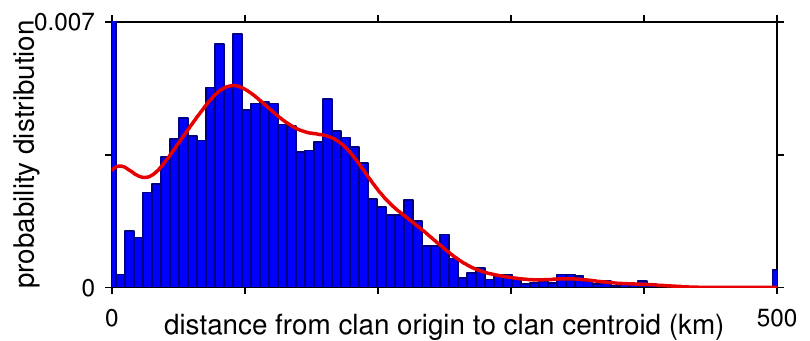}
\caption{Probability distribution for how far clans have moved in terms of the distance from the historical clan origin location to the clan centroid from 2000. We geographically identified the origin and centroid for 3\,900 clans among the 4\,303 clans in the 2000 census data. The rightmost bar corresponds to all distances of at least 500 km, and the solid curve is a kernel density estimate (from {\sc Matlab} R2011a's \texttt{ksdensity} function with default smoothing).
}
\label{movement_to_cm_fig}
\end{figure}

To obtain the coordinates of the clans' origins and the administrative regions, we wrote a {\sc Python} script that returns the latitude and longitude given a clan origin location's name. We used a {\sc Python} module for geocoding via Google Maps Application Programming Interface (API)~\cite{AGoogleMapAPI_licensing,AGoogleMapAPI_geocoding,AGoogleMapAPI_python}. For example, we were able to successfully retrieve 3\,900 clan origin locations out of the total of 4\,303 clans present in the 2000 census data (see Fig.~\ref{movement_to_cm_fig}).  We excluded the remaining 403 clan origin locations as erroneous because each of these cases has a distance of more than 1\,000 km between the identified origin location and the modern clan centroid.  (Such distances are much larger than the scale of the Korean peninsula.)

\begin{figure}[t]
\includegraphics[width=0.95\columnwidth]{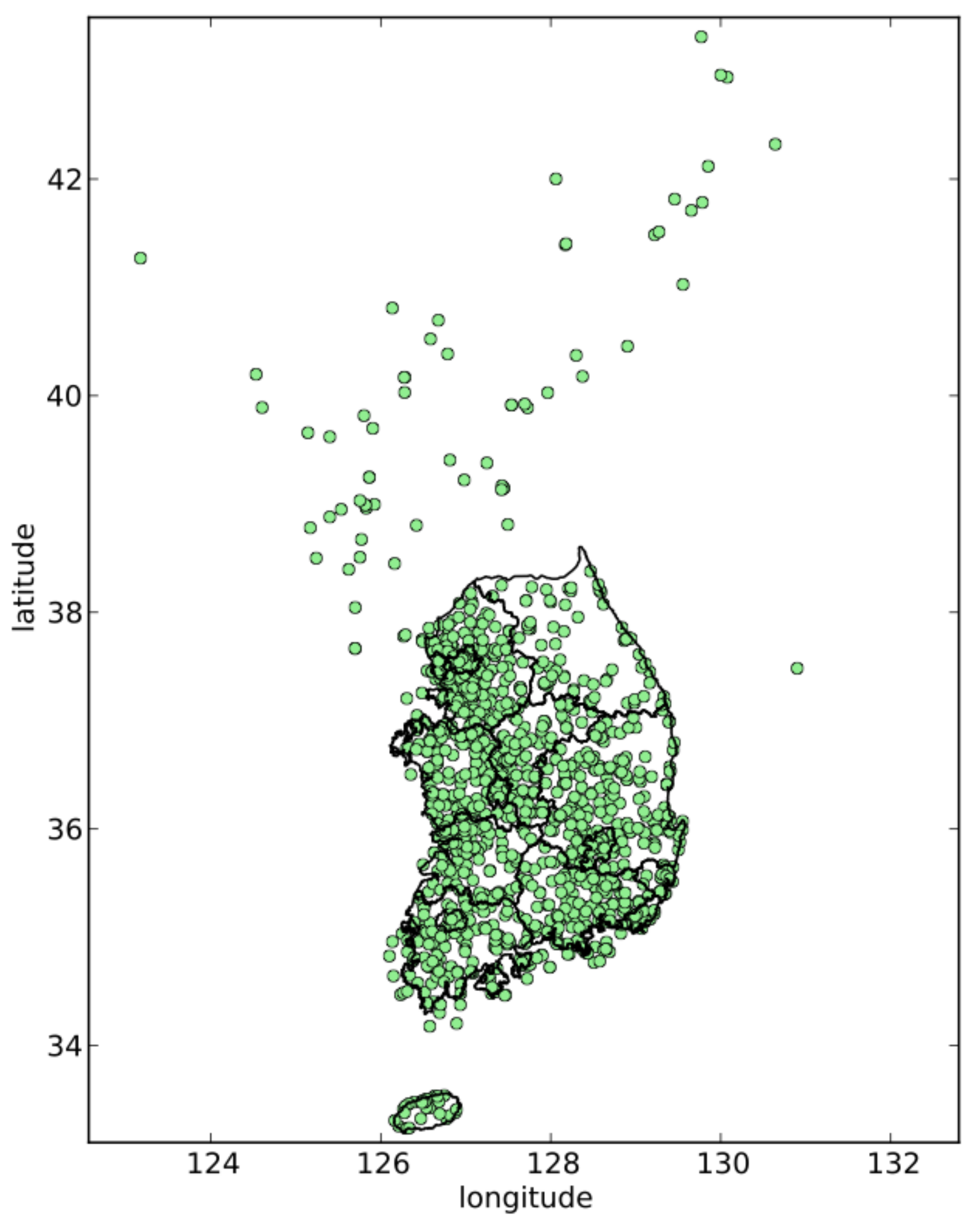}
\caption{Locations of clan origin names that we determined using the Google Maps API on top of the Korean map. We show administrative boundaries of South Korea
that correspond to the upper-level local autonomy (광역자치단체) composed of provinces (도), a special autonomous province (특별자치도),
a special city (특별시), 
and metropolitan cities (광역시)~\cite{Aadmin_divisions_SouthKorea}.
}
\label{origin_distribution}
\end{figure}

We confirmed by exhaustive checking that the modern administrative regions of South Korea are accurate.  (The first author, who is South Korean, manually checked all of the locations.)  However, as shown in Fig.~\ref{origin_distribution}, the clan origin locations are severely undersampled in the northern part of the Korean peninsula because of Google Maps' lack of information about North Korea. We hope to include more North Korean regions in future studies, and this might be possible because Google is adding details of North Korea to their mapping service~\cite{AGoogleMapNorthKorea}.

\begin{table}[t]
\caption{{\sc Python} code to obtain location coordinates in latitude and longitude from the Google Maps API. 
We set a delay of two seconds to avoid obtaining a message of Google Maps API's \texttt{OVER QUERY LIMIT}~\cite{Aover_query_limit}, as we are dealing with a large set of locations.
}
\begin{lstlisting}
from time import sleep
import sys

## https://bitbucket.org/xster/pygeocoder/wiki/Home
from pygeocoder import Geocoder

## get latlng from address
#TODO: edit this as required
address = 'Seoul, Korea'

try:
	sleep(2)

	results = Geocoder.geocode(address)
	(lat, lng) = results[0].coordinates

except:
	print 'error (addr2coord): ', sys.exc_info()[0]
	lat = -1
	lng = -1

print 'lat/lng : ', [lat, lng]

## retrieve associated address
try:
	sleep(2)

	results = Geocoder.reverse_geocode(lat, lng)
	retri_addr = str(results[0])

	print 'retri_addr: ', retri_addr

except:
	print 'error (coord2addr): ', sys.exc_info()[0]
\end{lstlisting}
\label{google_map_API_python_code}
\end{table}

In Table~\ref{google_map_API_python_code}, we present our {\sc Python} code using Google Maps API.  It requires {\sc pygeocoder} (we used version 1.1.4), which we downloaded on July 16, 2013~\cite{Apygeocoder}. The code returns coordinates in latitude and longitude, which can then be converted to metric units (see Appendix~\ref{sec:UTM}). 

\section{Universal Transverse Mercator Coordinates}
\label{sec:UTM}

All of the distance measures that we employ use the Universal Transverse Mercator (UTM) geographical coordinate system, which assigns a local two-dimensional Cartesian coordinate system to a given zone on the surface of the Earth~\cite{AUTM}. We use the UTM {\sc Python} module~\cite{AUTM_python}, which converts  $(\varphi,\lambda)$ coordinates for latitude ($\varphi$) and longitude ($\lambda$) to UTM coordinates (and vice versa), where the UTM standard revision used by this module is WGS84~\cite{AWGS84}. One can also convert from $(\varphi,\lambda)$ coordinates to UTM coordinates using Ref.~\cite{AUTM_coordinates_conversion}.

A point $(i_E, i_N)$ defined by UTM coordinates has two components: easting $i_E$ and northing $i_N$. For example, the mean UTM coordinates of our standardized regions are $i_E \approx 381.3$ and $i_N \approx 4\,017.7$, where the numbers are in units of kilometers from a reference point. The UTM scheme splits the Earth into 60 zones. Calculating distances between two points in different zones is complicated, in general, but the Korean peninsula lies entirely in zone 52~\cite{ASK_UTM_zone_52}, which simplifies the calculation considerably. For example, Seoul's (latitude, longitude) coordinates are $(\varphi,\lambda) \approx (37.58,127.00)$, and its UTM coordinates are $(i_E, i_N) \approx (323.4,  4\,161.5)$.  For our computations, we use formulas from Ref.~\cite{AUTMG}.  In a given zone, these formulas are accurate to within less than a meter.


\section{Czech Republic Surname Data}
\label{sec:CzechData}

The census data for the Czech Republic were derived from the 2009 Central Population Register (produced by the Czech Ministry of the Interior) by the authors of Ref.~\cite{czech_surname_pap}. The vast majority of the Czech Republic is within zone 33, although a small part of it is in zone 34~\cite{AUTM,ASK_UTM_zone_52}. As with the Korean clan origin locations, we used Google's API to roughly geolocate each of 206 Czech administrative regions by searching for the name of the administrative region followed by the string, ``, Czech Republic''.  We then converted all of the resulting latitudinal and longitudinal coordinates to UTM. (In this calculation, we assumed that all coordinates are in zone 33.)  

We use the surname concentration (i.e., surname population density) to define an ``anomaly'' that indicates the difference in value from what would be observed for an ``ergodic'' surname, which is well-mixed in a population. First, we obtain the centroid coordinates as in Eq.~\eqref{region_centroid}. The surname density anomaly is similar to what we defined for the Korean clans and is given by
\begin{equation}\label{anom2}
	\phi_i (k,t) =  c_i(k,t) - [m_i(t) / N(t) ] \rho(k,t) \, ,
\end{equation}
where $c_i(k,t)$ is the population density of surname $i$ in region $k$ at time $t$,
the quantity $m_i(t)$ is the total population of surname $i$ in all regions at time $t$,
the quantity $N(t)$ is the total population of all surnames in all regions at time $t$,
and $\rho(k,t)$ is the population density of all surnames in region $k$.  We use the same notational conventions as for Korean clans, so $\phi_i(k,t) = \phi_i[\mathbf{r}(k),t]$, $c_i(k,t) = c_i[\mathbf{r}(k),t]$, and $\rho(k,t) = \rho[\mathbf{r}(k),t]$.
It is necessary to introduce the anomaly (\ref{anom2}) 
because the total population is not conserved. Otherwise, we could simply take $\mathbf{J} \propto \nabla c_i$ as the flux of people with surname $i$. Instead, we take the flux to be
\begin{equation}
	\mathbf{J} \propto \nabla \phi_i\,.
\end{equation}	  

\section{Estimating Diffusion Constants}
\label{sec:estimating_diffusion_constants}

To estimate a diffusion constant for each Korean clan, we start with the known anomaly distribution based on 1985 census data. Using an initial guess for the diffusion constant $D_i$ (where $i$ indexes the clan), we integrate forward in time until 2000.  We then compare the numerical prediction to the known anomaly distribution based on 2000 census data using a single number: the relative error, which we define as the sum of squared deviations divided by the sum of squared anomalies from 2000. 

After we calculate the relative error, we can adaptively change the ``guess'' for $D_i$ and repeat the above process until we find the optimal $D_i$ values. In practice, we use {\sc Matlab}'s built-in numerical optimization routine \texttt{fminbnd}~\cite{Afminbnd}, which implements a Nelder-Mead downhill simplex search.

\subsection{Some subtleties}

Because the census data are irregular, we first interpolate them to a regular grid before numerically integrating the diffusion equation.  The grid that we use covers the UTM-zone-52 rectangular region from 245 to 545 km easting and from $3\,800$ to $4\,250$ km northing with a uniform 2.5 km spacing between grid points. (We exclude Jeju Island, which is distant from mainland Korea and is located south of the mainland.)  We employ a standard five-point stencil to approximate the Laplacian operator in space and integrate in time with a 4th- or 5th-order adaptive Runge-Kutta scheme.  We impose Neumann conditions at the boundaries.

Because the numerical integration is unstable for negative values of $D_i$, we restrict $D_i$ to be positive.  We test for the possibility of ``negative diffusion'' by repeating the entire optimization procedure after interchanging the 1985 and 2000 data sets; that is, we start from 2000 and integrate backwards in time with a positive diffusion constant.


\subsection{Testing against a null hypothesis}

Because our hypothesis of clan diffusion is somewhat speculative, it is important to test it against a basic null hypothesis.  We take the null hypothesis to be a model in which clans do indeed diffuse but in which clan affiliation plays no role. Members simply diffuse in accordance with the local population density $\rho$.  Therefore, 
\begin{equation} \label{diffusion_eq_null}
  	\frac{\partial \phi_i}{\partial t} = D_i \, \nabla^2 \rho\,.
\end{equation}
We accept the null model as preferable to the clan-diffusion model whenever it yields a lower relative error using its best-fit $D_i$.


\subsection{Numerical results}

The results of our computational examination of diffusion are as follows.  When we include all $3\,481$ clans for which data are available in both the 1985 and 2000 censuses, we obtain a mean diffusion constant $\bar{D} = \langle D_i \rangle$ (where we average over the clans) of $\bar{D} \approx 2.91$ and a standard deviation of $\sigma _D \approx 10.4$. When we remove ``small'' clans (i.e., those with fewer than $2\,000$ members in the year 2000), the 707 remaining clans have a mean diffusion constant of $\bar{D} \approx 0.79$ and a standard deviation of $\sigma_D \approx 4.8$.  When we also remove ``ergodic clans'' (by eliminating the 75\% with the largest year-2000 radii of gyration), the 182 remaining clans have a mean diffusion constant of $\bar{D} \approx 1.3$ and a standard deviation of $\sigma_D \approx 2.0$.

When we use all $3\,481$ clans, our computations favor the clan-diffusion model over the null model in about $84$\% of the cases. Additionally, about $64$\% of all clans have both positive best-fit diffusion coefficients \textit{and} mobility patterns that are explained better by the clan-diffusion model.

When we exclude both small and ergodic clans, our computations favor the clan-diffusion model over the null model in about $81$\% of the cases (i.e., for 148 clans). Moreover, 78\% of all clans have both positive best-fit diffusion constants \textit{and} mobility patterns that are explained better by the clan-diffusion model than by the null model.

In Fig.~\ref{diffusion_constant_distribution}, we show a histogram of diffusion constants for the subset of clans for which the clan-diffusion model appears to be valid. These 148 clans are nonergodic, have a positive best-fit diffusion constant, and are fit better by the clan-diffusion model than by the null model. They have a mean diffusion constant of $\bar{D} \approx 1.6$ and a standard deviation of $\sigma_D \approx 2.1$.

\section{Construction of a Population-Flow Network between Regions}
\label{sec:centrality_of_seoul}

\begin{figure}[t]
\includegraphics[width=0.9\columnwidth]{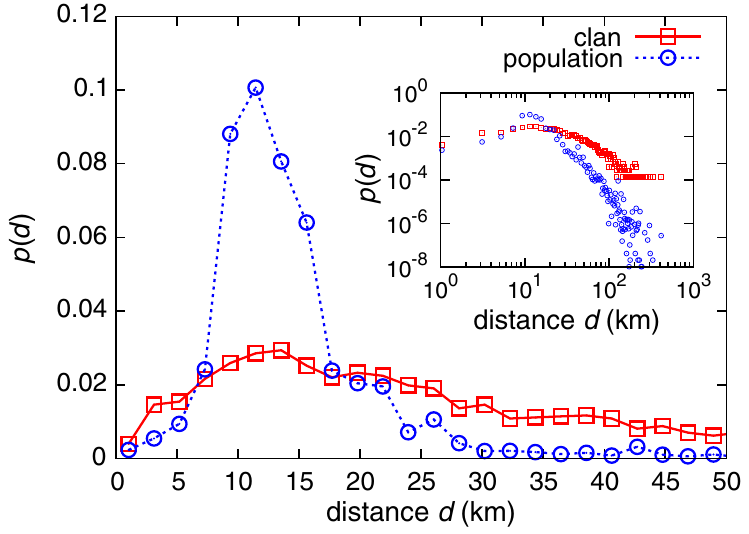} 
\caption{
Distribution of clan-centroid movement between 1985 and 2000, which we compute using $\lVert \mathbf{r}_{i,\textrm{C}}^{\textrm{pop}} (t=1985) - \mathbf{r}_{i,\textrm{C}}^{\textrm{pop}} (t=2000) \rVert$ from Eq.~\eqref{eq:C_formula_population}.  (The norm $\lVert \cdot \rVert$ is the Euclidean norm.) The curve marked by squares weighs each clan equally, and the curve marked by circles weighs each clan by its population.  We indicate distance in units of populations in 2000. (The 1985 data give a very similar distribution.)  As one can see in the inset (for which we use a doubly logarithmic scale), the maximum movement distance is larger than $400$ km. However, as the main panel illustrates, most clans moved considerably shorter distances.
}
\label{distance_distribution_real}
\end{figure}

Although it is impossible to track the movement of individual people from the census data (because the data do not include such information),
it is possible to construct a rough estimate of the population flow between a pair of regions by considering the movement of clans (i.e., of the smallest demographic unit that it is possible to resolve with our data) 
between 1985 and 2000. For each clan $i$ of the 3481 clans that appear in both the 1985 and 2000 census data, we define the population centroid [note the contrast with the clan \emph{anomaly} centroid from Eqs.~\eqref{C_formula} and \eqref{normalization_constant}] as
\begin{equation}\label{eq:C_formula_population}
  	\mathbf{r}_{i,\textrm{C}}^{\textrm{pop}} (t) = \frac{1}{c_{i,\textrm{tot}}(t)} \sum_k \mathbf{r}(k) c_i(k,t) A_k \,, \\
\end{equation}
where the normalization constant is
\begin{equation}\label{eq:normalization_constant_population}
	c_{i,\textrm{tot}} (t) = \sum_k c_i (k,t) A_k \equiv \sum_k N_i(k,t)\,.
\end{equation}
Recall that $k$ indexes the region in Korea, ${\bf r}(k) = [x(k),y(k)]$ gives the coordinate of that region's centroid, $A_k$ is its area, and $c_i(k,t)$ is the population density of clan $i$ in region $k$ at time $t$. Additionally, recall that $N_i(k,t)=c_i(k,t)A_k$ is the population of clan $i$ in region $k$ at time $t$ (see Sec.~\ref{sec:human_diffusion}).

We considered approximating the movement of each clan $i$ by 
\begin{equation*}
	 \mathbf{r}_{i,\textrm{C}}^{\textrm{pop}} (t=2000) - \mathbf{r}_{i,\textrm{C}}^{\textrm{pop}} (t=1985)\,,
\end{equation*}
but this would entail treating an entire clan population as a ``point mass'', so it neglects valuable information from the spatial variation (as illustrated by our calculations of clan ergodicity). In addition, as we show in Fig.~\ref{distance_distribution_real}, the amount of movement for the majority of clans is too small to proceed further with such an approach. (One can also infer the small scale of clan movements from Fig.~\ref{diffusion_constant_distribution}.)

\begin{figure*}[t]
\includegraphics[width=\textwidth]{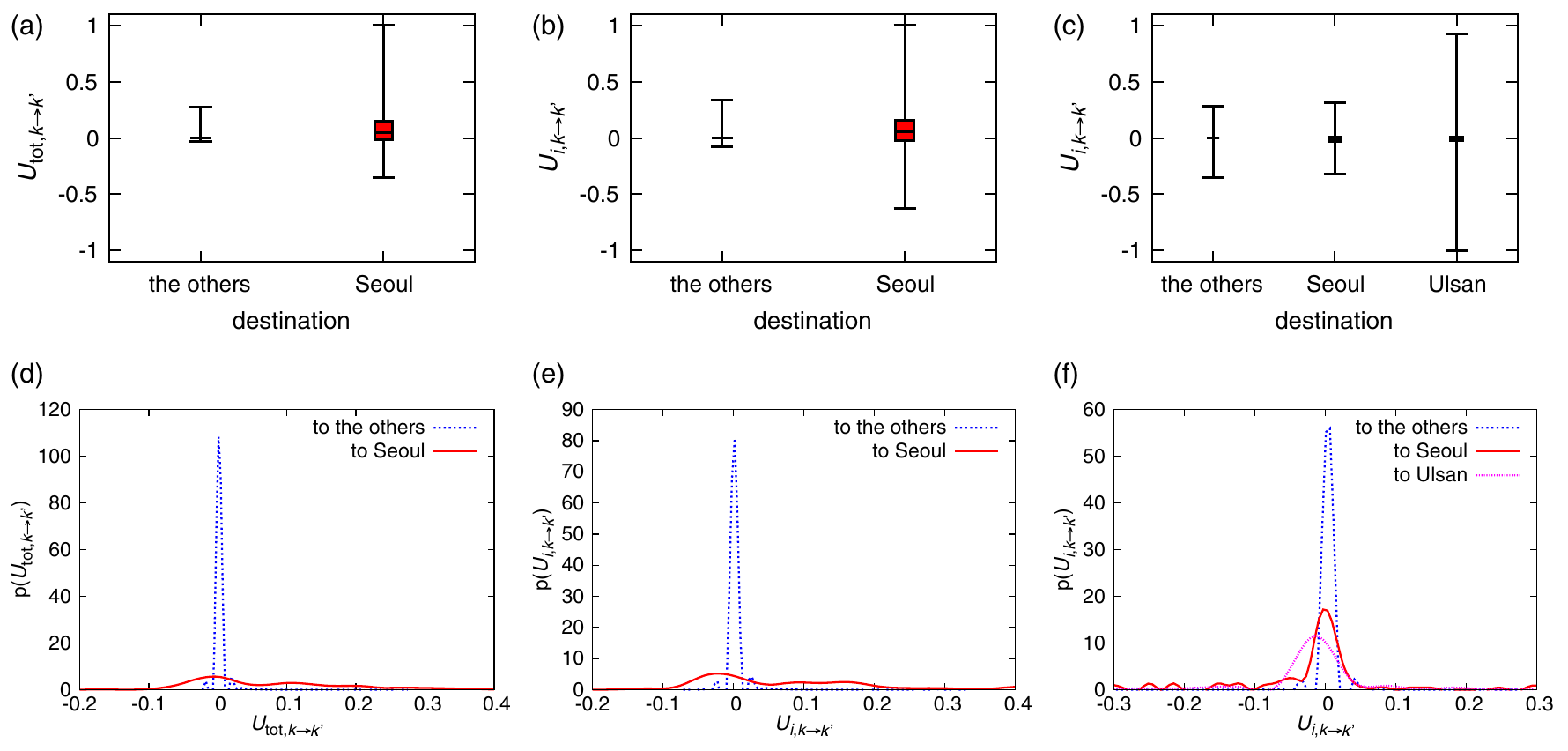}
\caption{We show box plots for the distributions of (a) $U_{\textrm{tot},k \to k'}$ for total population, (b) $U_{i,k \to k'}$ for Kim from Gimhae, and (c) $U_{i,k \to k'}$ for Lee from Hakseong, where the boxes indicate quartile values and the whiskers give the minimum and maximum values.
We also show the distributions for (d) $U_{\textrm{tot},k \to k'}$ for total population, (e) $U_{i,k \to k'}$ for Kim from Gimhae, and (f) $U_{i,k \to k'}$ for Lee from Hakseong.  In all cases, we apply \texttt{cspline} smoothing in \texttt{gnuplot} to the histograms using $50$ equally-sized bins.
}
\label{W_boxplot}
\end{figure*}

As an alternative that avoids the undesirable point-mass approximation, we attempt to infer the flow of a clan between two regions from changes in population ratios.
 Let each of the $199$ standardized administrative regions of South Korea (see Appendix~\ref{sec:standardized_regions} for details) be individual nodes of a population-flow network~\cite{NetworkReview} between 1985 and 2000. To examine the central nature of Seoul in such a network, we merge the $17$ regions that correspond to different parts of Seoul into a single node that we call ``Seoul''. The resulting population-flow network has $183$ nodes.
The raw change in relative populations between regions $k$ and $k'$ in our census data is 
\begin{equation}   \label{eq:relative_fraction_difference}
 	\tilde{W}_{i,k \to k'} = \frac{N_i (k',t=2000)}{N_i (k,t=2000)} - \frac{N_i (k',t=1985)}{N_i (k,t=1985)} \,,
\end{equation}
where 
we exclude the regions with $N_i(k,t=1985) = 0$ or $N_i(k,t=2000) = 0$ to avoid singularities.  We then define the normalized relative population change as
\begin{equation}   \label{eq:normalized_relative_fraction_difference}
  	U_{i,k \to k'} = \frac{\tilde{W}_{i,k \to k'}}{\max_{q \to q'}\left\{ \left| \tilde{W}_{i,q \to q'} \right| \right\} } \,,
\end{equation}
so that $U_{i,k \to k'} \in [-1,1]$.

The quantity $U_{i,k \to k'}$
is a proxy for a more finely-grained quantity, which we denote by $W_{i,k \to k'}$, that describes the real population flow of clan $i$ from region $k$ to region $k'$ (so $k \to k'$ is a directed edge) and would be desirable to construct from data for flows of individuals.  The smallest units in the census data are clans, so we need to estimate population flows from them as our basic units.
We thus instead calculate $\tilde{W}_{i,k \to k'}$ and use the normalized relative population changes $U_{i,k \to k'}$ in Eq.~\eqref{eq:normalized_relative_fraction_difference} as the adjacency-matrix elements of a weighted and directed population-flow network.

Our proxy is not guaranteed to be ``correct'', but it has several properties that are consistent with reasonable flow behavior: (i) If the ratio $N_1 / N_2$ of clan populations (in regions $k'=1,k=2$) does not change from 1985 to 2000, then both the proxy flow $U_{i,2 \to 1}$ and the
inferred flow $W_{i,2 \to 1}$ from region $2$ to region $1$ are equal to $0$; 
(ii) if the ratio $N_1 / N_2$ increases from 1985 to 2000, then the proxy and inferred flows from region $2$ to region $1$ are both positive; and (iii)
if the ratio $N_1 / N_2$ decreases from 1985 to 2000, then the proxy and inferred flows from region $2$ to region $1$ are both negative.
Naturally, both the proxy flow and the inferred flow are asymmetric (so $W_{i,2 \to 1} \neq -W_{i,1 \to 2}$ in general).

As a downside, uniform population growth biases the proxy calculation slightly in favor of flow towards regions with smaller populations.
 Additionally, the proxy cannot capture circulating flows and is unlikely to do a good job when flow is strongly multipolar (i.e., if more than one area attracts a significant amount of flow).  When flow is mostly between Seoul and other regions, we call it ``unipolar''.

Using Eq.~\eqref{eq:normalized_relative_fraction_difference}, we define a population-flow network for each clan $i$. We include all clans by using the entire population density of region $k$ in year $t$. In other words, we calculate $N_{\textrm{tot}} (k,t) = \sum_i N_i (k,t)$
and obtain a raw total population-flow network, which has corresponding adjacency-matrix elements 
\begin{equation}  \label{eq:total_relative_fraction_difference}
	  \tilde{W}_{\textrm{tot},k \to k'} = \frac{N_{\textrm{tot}} (k',t=2000)}{N_{\textrm{tot}} (k,t=2000)} - \frac{N_{\textrm{tot}} (k',t=1985)}{N_{\textrm{tot}} (k,t=1985)} \, .
\end{equation}
The adjacency-matrix elements for the associated normalized, relative total population-flow network are
\begin{equation}  \label{eq:total_normalized_relative_fraction_difference}
  	U_{\textrm{tot},k \to k'} = \frac{\tilde{W}_{\textrm{tot},k \to k'}}{\max_{q \to q'}\left\{ \left| \tilde{W}_{\textrm{tot},q \to q'} \right| \right\} } \,.
\end{equation}
Our normalization guarantees that $U_{\textrm{tot},k \to k'} \in [-1,1]$.

In Fig.~\ref{W_boxplot}(a), we show box plots for the distribution of $U_{\textrm{tot},k \to k'}$ using all pairs $k,k'$.  We also show box plots for the distributions of $U_{i,k \to k'}$ for the clans Kim from Gimhae and Lee from Hakseong.  We show flows to Seoul separately from flows to other regions.
Note that the values of $U_{\textrm{tot},k \to k'}$ are distributed much more broadly for flows to Seoul than for flows to other regions, even though there are many more adjacency-matrix elements for the latter ($182$ for flows to Seoul and $33\,124$ for flows to other regions).  One can also observe this feature in the distribution shapes themselves [see Figs.~\ref{W_boxplot}(d)--\ref{W_boxplot}(f)].

\begin{figure*}[t]
\begin{tabular}{lll}
{\sffamily (a)} & {\sffamily (b)} & {\sffamily (c)} \\
\includegraphics[width=0.25\textwidth]{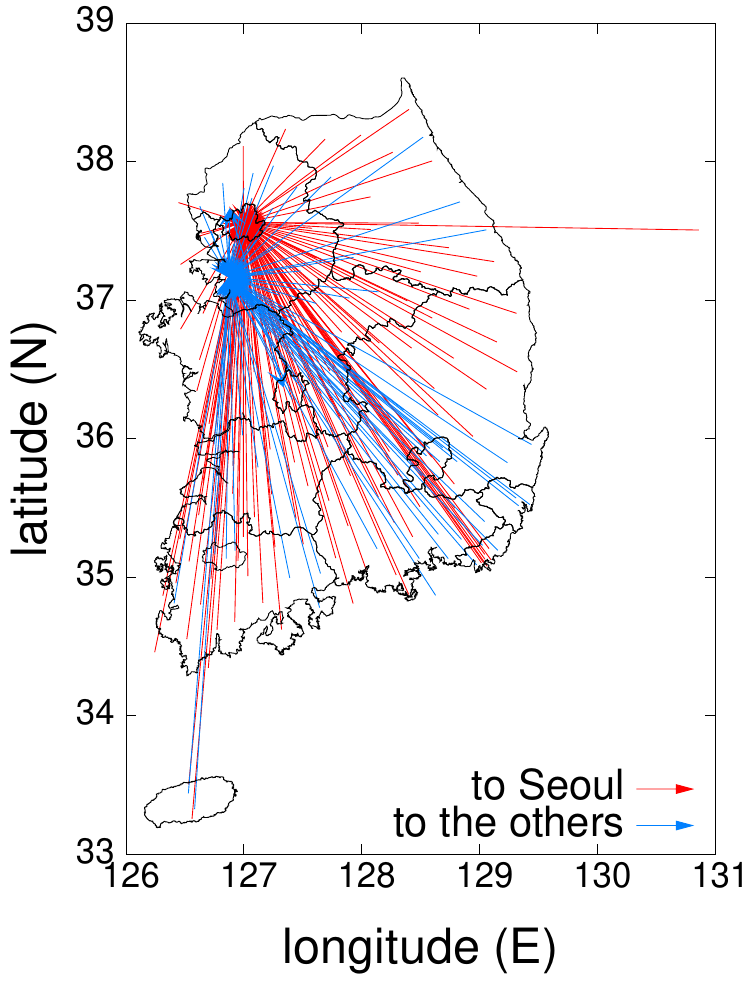} & 
\includegraphics[width=0.25\textwidth]{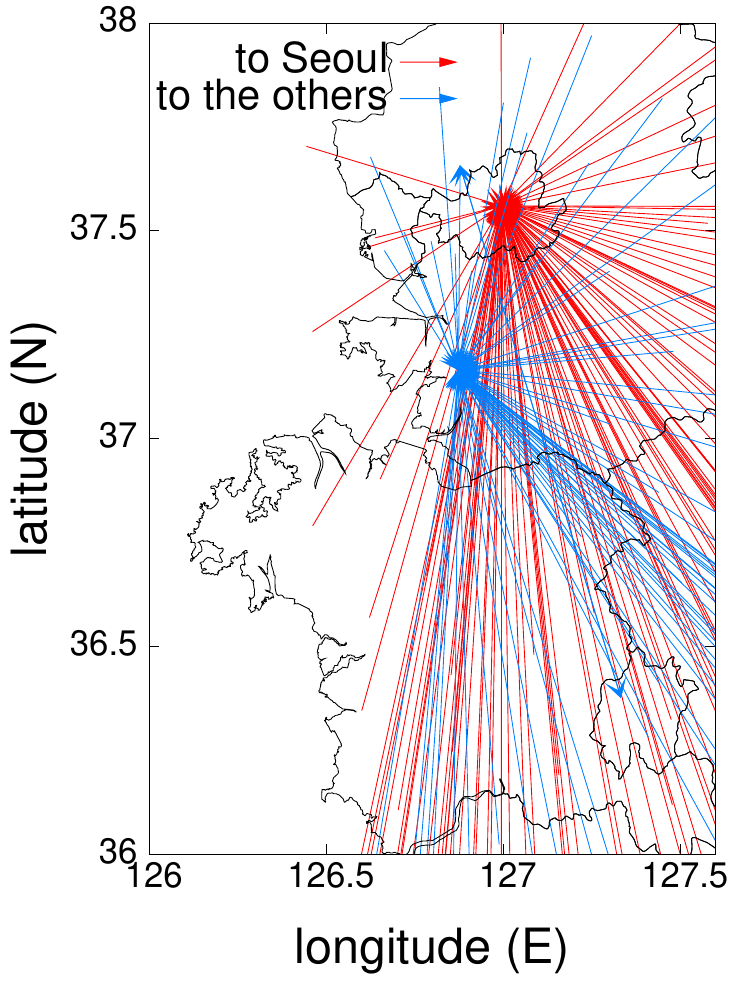} 
 & \includegraphics[width=0.25\textwidth]{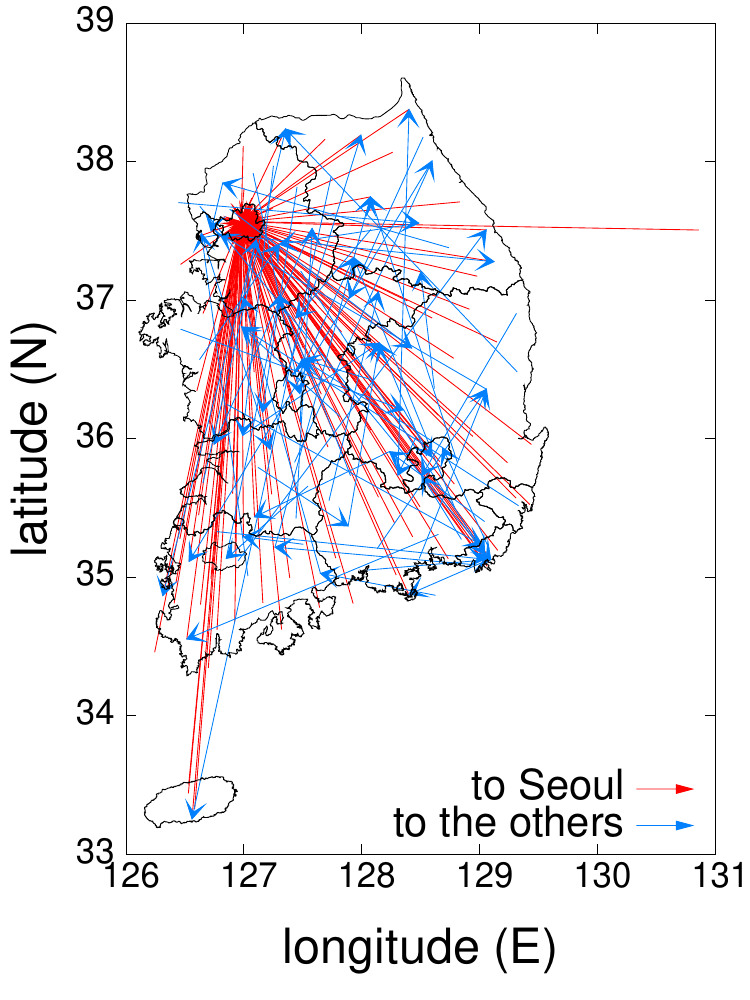} \\
\end{tabular}
\caption{
(a) Maximum relatedness subnetwork (MRS)~\cite{GoogleCorr} of the combined population-flow network in Eq.~\ref{eq:total_normalized_relative_fraction_difference} for all clans, (b) a magnified portion of this MRS that includes all regions with nonzero in-degrees, and (c) an instance of a model (inspired by the one in Ref.~\cite{Morris2012}) of a rewired version of a monocentric network (with Seoul as the center) with a rewiring probability of $p = 0.4$. (See the discussion in the text.) Red lines show the directed edges towards Seoul, and blue lines show the directed edges towards other regions.
}
\label{MRS_Korea}
\end{figure*}

\begin{table*}[t]
\caption{
List of regions (which we name based on the current administrative regions) with nonzero values of in-degree in our MRSs.  
}
\begin{ruledtabular}
\begin{tabular}{llrr}
Clan: figure & Region & In-degree & Percentage of total in-degree \\
\hline
All clans: & Seoul (서울) & $109$ & $59.9\%$ \\
\cline{2-4}
Figs.~\ref{MRS_Korea}(a) and (b) & Hwaseong (화성) + Ansan (안산) + Osan (오산) & $70$ & $38.7\%$ \\
 & + Siheung (시흥) + Gwacheon (과천) + Gunpo (군포)  & & \\
 & + Euiwang (의왕) of Gyeonggi Province (경기도) & & \\
\cline{2-4}
 & Goyang (고양) of Gyeonggi Province (경기도) & $2$ & $1.1\%$ \\
\cline{2-4}
 & Yuseong-gu (유성구) of Daejeon (대전) & $1$ & $0.5\%$ \\
\hline
Kim from Gimhae & Seoul (서울) & $106$ & $58.2\%$ \\
\cline{2-4}
(김해 김): & Hwaseong (화성) + Ansan (안산) + Osan (오산) & $74$ & $40.7\%$ \\
Fig.~\ref{MRS_Korea_GimhaeKim_HakseongLee}(a) & + Siheung (시흥) + Gwacheon (과천) + Gunpo (군포) & & \\
 & + Euiwang (의왕) of Gyeonggi Province (경기도) & & \\
\cline{2-4}
 & Yuseong-gu (유성구) of Daejeon (대전) & $1$ & $0.5\%$ \\
\cline{2-4}
 & Daedeok-gu (대덕구) of Daejeon (대전) & $1$ & $0.5\%$ \\
\hline
Lee from Hakseong & Nam-gu (남구) of Ulsan (울산) & $52$ & $31.0\%$ \\
\cline{2-4}
(학성 이): & Yongin (용인) of Gyeonggi Province (경기도) & $41$ & $24.4\%$ \\
\cline{2-4}
Fig.~\ref{MRS_Korea_GimhaeKim_HakseongLee}(b) & Goyang (고양) of Gyeonggi Province (경기도) & $37$ & $22.0\%$ \\
\cline{2-4}
 & Jung-gu (중구) + Buk-gu (북구) & $28$ & $16.7\%$ \\
 & + Ulju-gun (울주군) of Ulsan (울산) & & \\
\cline{2-4}
 & Buk-gu (북구) + Gangseo-gu (강서구) & $8$ & $4.8\%$ \\
 & + Sasang-gu (사상구) of Busan (부산) & & \\
\cline{2-4}
 & Yeongi (연기) of Chungcheongnamdo (충청남도) & $2$ & $1.2\%$ \\
\end{tabular}
\end{ruledtabular}
\label{incoming_region_list} 
\end{table*}

One simple but intuitive way to check the centrality of Seoul is to extract a maximum relatedness subnetwork (MRS) \cite{GoogleCorr} from each population-flow network.  We construct a MRS as follows. For each node, we examine the weight of each of its edges and keep only the single directed edge with maximum weight. (When there are ties, we keep all edges that have the maximum weight.) We exclude out-edges from Seoul for the MRS in order to focus on the movement from other regions to Seoul.  We will later compare the MRS to a null-model network that also disallows out-edges from the central node.
In Figs.~\ref{MRS_Korea}(a) and \ref{MRS_Korea}(b), we show the MRS from the adjacency matrix with elements $U_{\textrm{tot},k \to k'}$.
 The central role of Seoul is apparent. As we indicate in Table~\ref{incoming_region_list}, Seoul's in-degree in this MRS is $109$.  This constitutes nearly $60\%$ of the MRS edges and is consistent with the rapid growth of the Seoul area that we illustrate in Fig.~\ref{cartogram} in Appendix~\ref{sec:other_results}. 

We model the population flow using a simple rewired-network model inspired by the model in Ref.~\cite{Morris2012}.  We start with a ``monocentric'' network, with Seoul as the central node, in which all directed edges start in some region (aside from the center) and terminate at Seoul.  We then rewire each edge with independent, uniform probability $p$ by randomly choosing the terminal end.   
Each network that we construct in this way has one directed edge for each node aside from the central one, so we can use an ensemble of such networks as a null model for our MRSs.

As we indicate in Table~\ref{incoming_region_list}, the edges in the MRSs are distributed rather heterogeneously among the regions. For example, the region in Gyeonggi Province (which has the second-largest in-degree) has about $39\%$ of the edges for the MRS that we constructed using all clans.  When constructing null-model networks, we use a rewiring probability of $p = 0.4$ to ensure that about $60\%$ of the directed edges terminate in Seoul on average (as suggested by the data when considering all clans). The null-model network ensemble generated from the rewiring process has a binomial (or approximately Poisson, as the MRS is rather sparse) in-degree distribution as a result of the given fraction $p$ of edges that are redirected uniformly at random except for the central node (i.e., Seoul)~\cite{NetworkReview}. Therefore, the emergence of a second-largest hub comparable
in size to the largest hub (Seoul) is extremely unlikely. We illustrate one instance of such a rewired network in Fig.~\ref{MRS_Korea}(c), and the MRS for all clans that we constructed from empirical data differs significantly from the null-model network (see Table~\ref{incoming_region_list} as well).

\begin{figure}[t]
\begin{tabular}{ll}
{\sffamily (a)} & {\sffamily (b)} \\
\includegraphics[width=0.5\columnwidth]{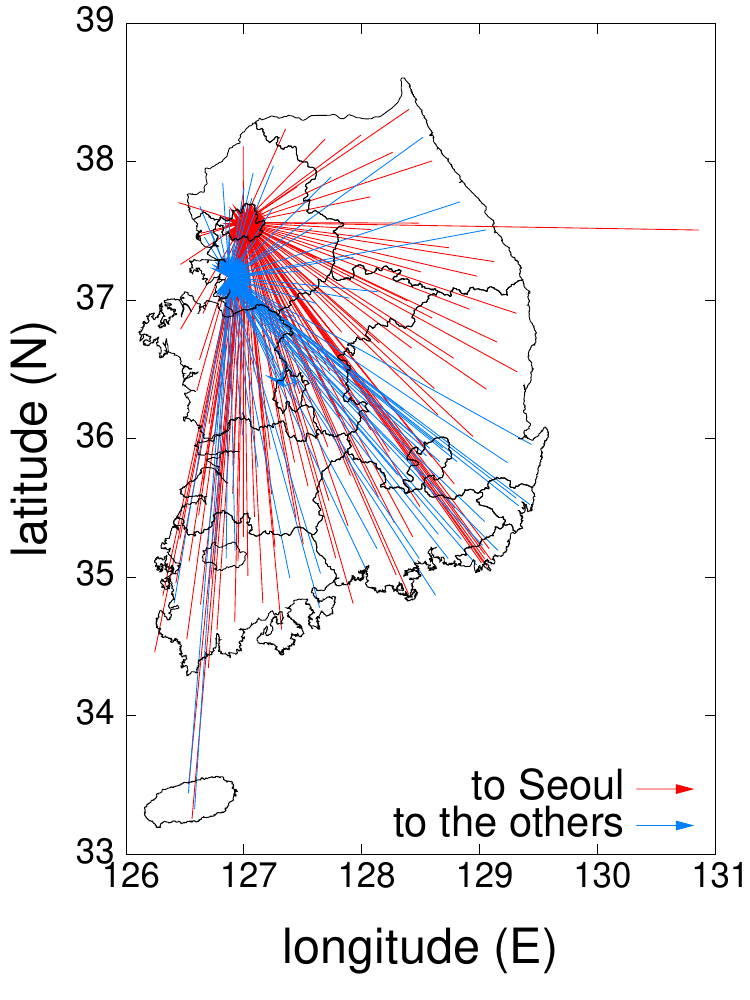} & \includegraphics[width=0.5\columnwidth]{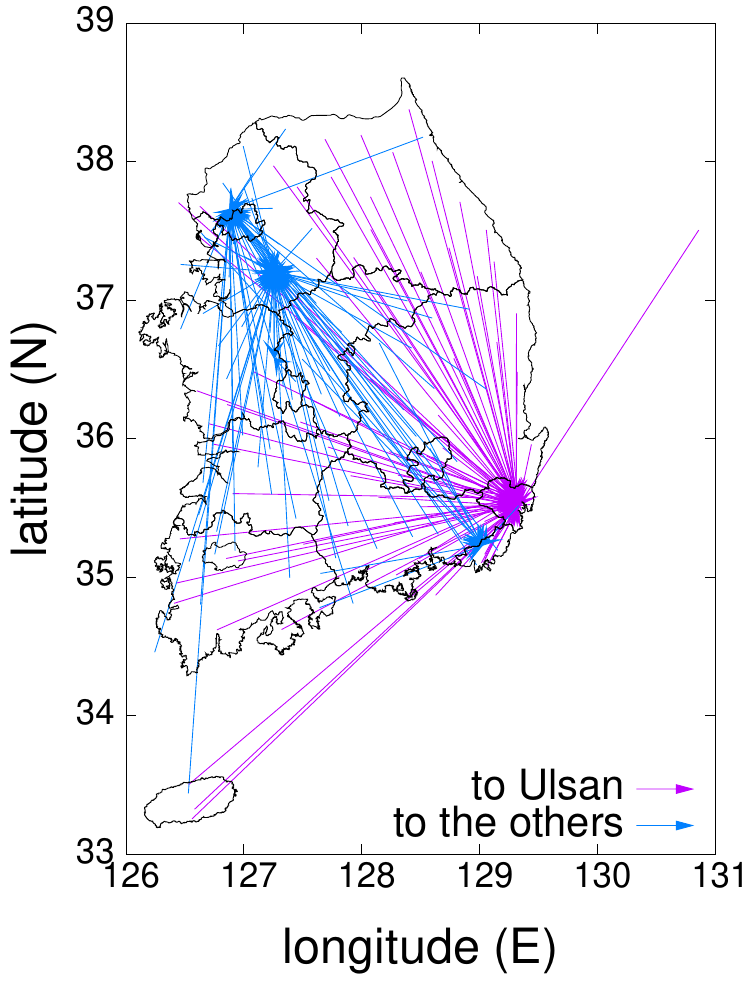} \\
\end{tabular}
\caption{
(a) MRS~\cite{GoogleCorr} of the population-flow network for (a) Kim from Gimhae and (b) Lee from Hakseong.  In panel (a), red lines indicate directed edges towards Seoul, and blue lines indicate directed edges towards other regions. In panel (b), we show directed edges towards Ulsan in purple and directed edges towards other regions in blue.
}
\label{MRS_Korea_GimhaeKim_HakseongLee}
\end{figure}

It is also instructive to examine the population-flow networks for individual clans.  As with prior discussions, we will use Kim from Gimhae as an example of an ergodic clam and Lee from Hakseong as an example of a nonergodic clan (see Fig.~\ref{GimhaeKim_vs_HakseongLee}).

When we consider the population-flow network for the clan Kim from Gimhae [by using $N_i(k,t)$ with $i$ corresponding to Kim from Gimhae in Eq.~\eqref{eq:relative_fraction_difference}], we obtain a qualitatively similar result---namely, an abundance of edges terminating in Seoul---to what we obtained when using all clans. 
See Figs.~\ref{W_boxplot}(b) and \ref{MRS_Korea_GimhaeKim_HakseongLee}(a), and Table~\ref{incoming_region_list}. By contrast, we find that two different locations ``attract'' the population for Lee from Hakseong.  Following the general trend in the population, one area is the Gyeonggi Province in the northwestern part of South Korea that surrounds the Seoul area. (The name ``Gyeonggi'' means ``the area surrounding capital'' in Korean and it is often construed to be essentially an ``extended Seoul''.)
The other area is Ulsan/Busan in the southeastern part of South Korea (where the clan origin is located).
See Figs.~\ref{W_boxplot}(c) and \ref{MRS_Korea_GimhaeKim_HakseongLee}(b), and Table~\ref{incoming_region_list}.
As one can see from Fig.~\ref{W_boxplot}(c), the Seoul region is not special for this clan. 
Therefore, we see that this young, nonergodic clan has a different mobility pattern from the stabilized, ergodic clans that follow the general trend in population flow.

\section{Other Results}
\label{sec:other_results}

\begin{figure}[t]
  \includegraphics[width=0.9\columnwidth]{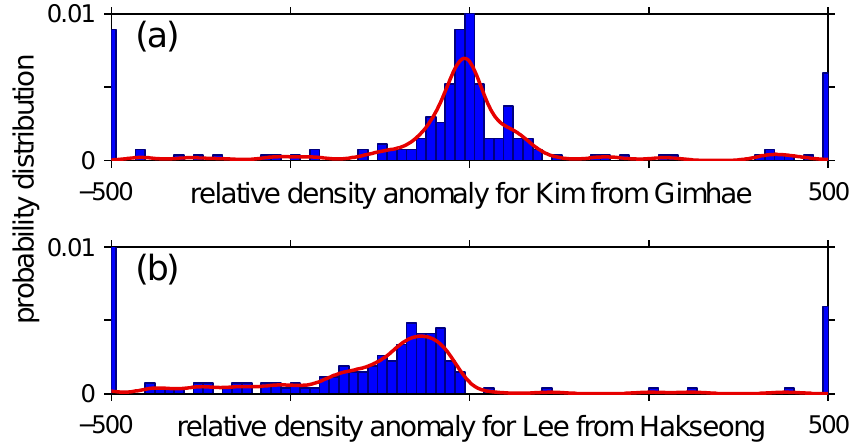} 
  \caption{Distributions of clan-density anomalies $\{ \phi_i \}$ in 2000 over the 199 standardized administrative regions for (a) Kim from Gimhae and (b) Lee from Hakseong. The leftmost and rightmost peaks correspond, respectively, to all values less than or equal to $-500$ and all values greater than or equal to $500$. Solid curves are kernel density estimates (from {\sc Matlab} R2011a's \texttt{ksdensity} function with a Gaussian smoothing kernel of width 20).
}
  \label{phi_distribution_for_Kim_and_Lee}
\end{figure}

\begin{figure}[b]
  \includegraphics[width=0.85\columnwidth]{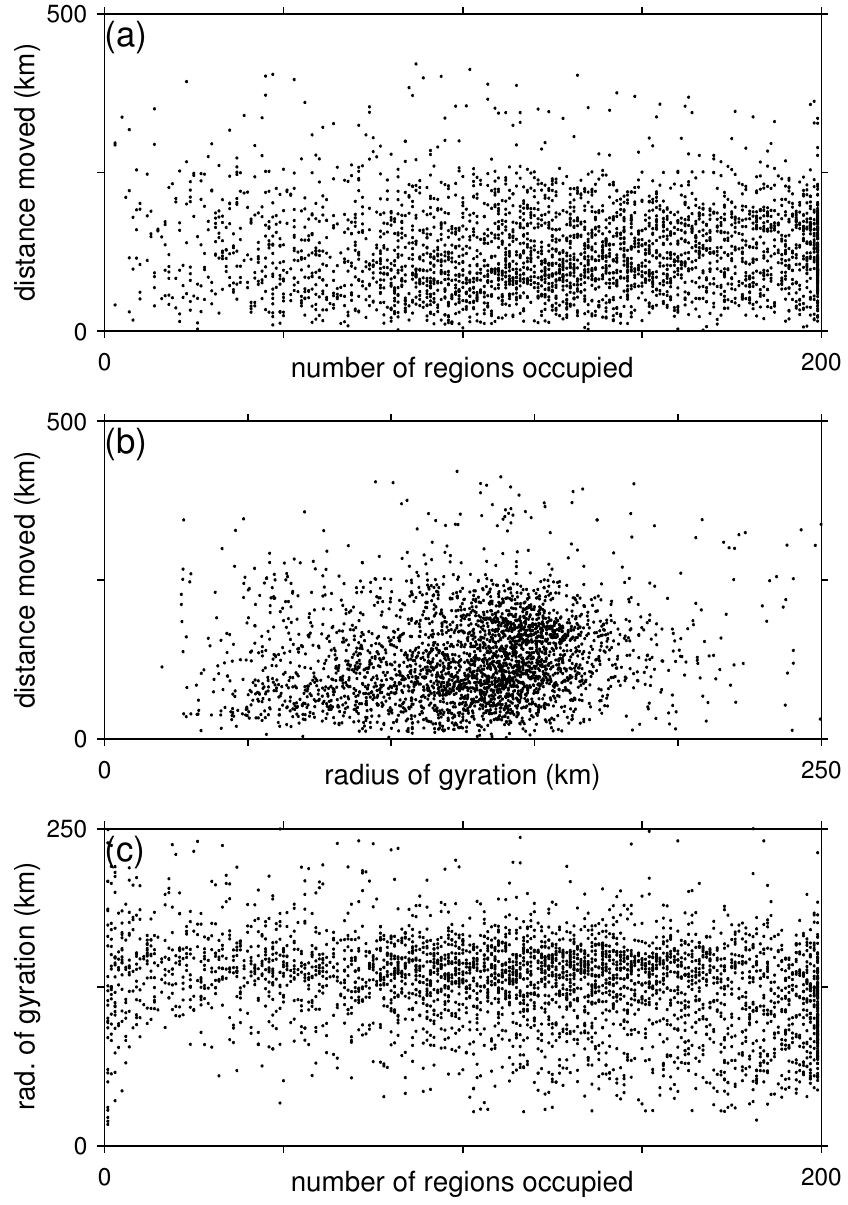}
  \caption{
Scatter plots of (a) distance between clan origin location and population centroid versus number of administrative regions, (b) distance between clan origin location and population centroid versus radius of gyration, and (c) radius of gyration versus number of administrative regions. (In this figure, we consider the $3\,481$ clans for which data are available in both the 1985 and 2000 censuses; We were able to determine the origin location for $3\,120$ of these clans.) The corresponding Pearson correlation values are (a) $r \approx -0.0094$ (from the $3\,120$ clans for which we know the origin location; the $p$-value is $p \approx 0.60$), (b) $r \approx 0.14$ (from the $3\,120$ clans for which we know the origin location; the $p$-value is $p \approx 1.1 \times 10^{-15}$), and (c) $r \approx -0.26$ (from the $3\,481$ clans; the $p$-value is $p \approx 1.8 \times 10^{-53}$).  Note that correlations over limited ranges can be different and significant. For example, in panel (c), the two diagnostics for ergodicity are significantly positively correlated when $r_g<50$ km ($r \approx 0.39$, $p \approx 2.3 \times 10^{-4}$). 
}
  \label{distance_num_admin_RG}
\end{figure}

\begin{figure}[b]
  \includegraphics[width=0.95\columnwidth]{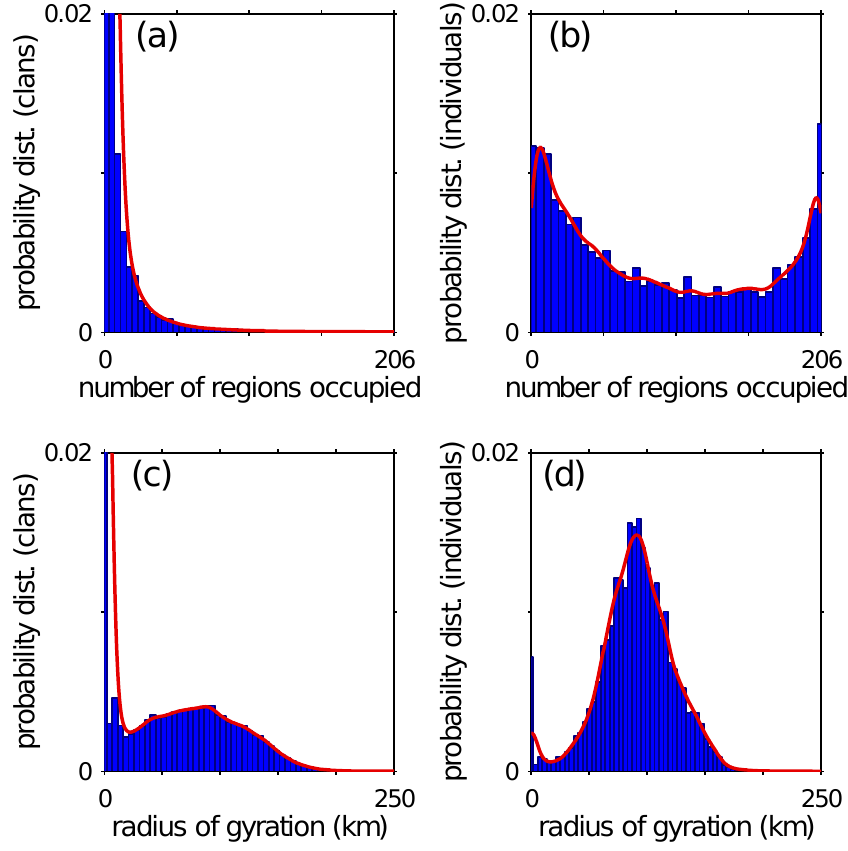} 
  \caption{(a) Probability distribution of the number of different administrative regions occupied a Czech family name in 2009. Note that the two leftmost bars have heights of $0.17$ and $0.03$ (with $0.17$ at the farthest left), but we have truncated them for visual presentation. (These data were initially analyzed in Ref.~\cite{czech_surname_pap}.) 
(b) Probability distribution of the number of different administrative regions occupied by the clan of a Czech individual selected uniformly at random in 2009. The difference between this panel and the previous one arises from the fact that clans with larger populations tend to occupy more administrative regions. [That is, we select a clan uniformly at random in panel (a), but we select an individual uniformly at random in panel (b).] (c) Probability distribution of radii of gyration (in km) of Czech family names in 2009. Note that the leftmost bar has a height of $0.11$, but we have truncated it for visual presentation.
(d) Probability distribution of radii of gyration (in km) of Czech family names of a Czech individual selected uniformly at random in 2009.  The difference between this panel and the previous one arises from the fact that clans with larger populations tend to occupy more administrative regions.
Observe that the distributions in panels (a) and (b) are starkly different from the distributions in panels (a) and (b) from Fig.~\ref{occu_num_diff_admin}. 
Solid curves are kernel density estimates (from {\sc Matlab} R2011a's \texttt{ksdensity} function with a Gaussian smoothing kernel of width 5).}
  \label{occu_num_diff_admin_Czech}
\end{figure}

\begin{figure}[t]
  \includegraphics[width=0.95\columnwidth]{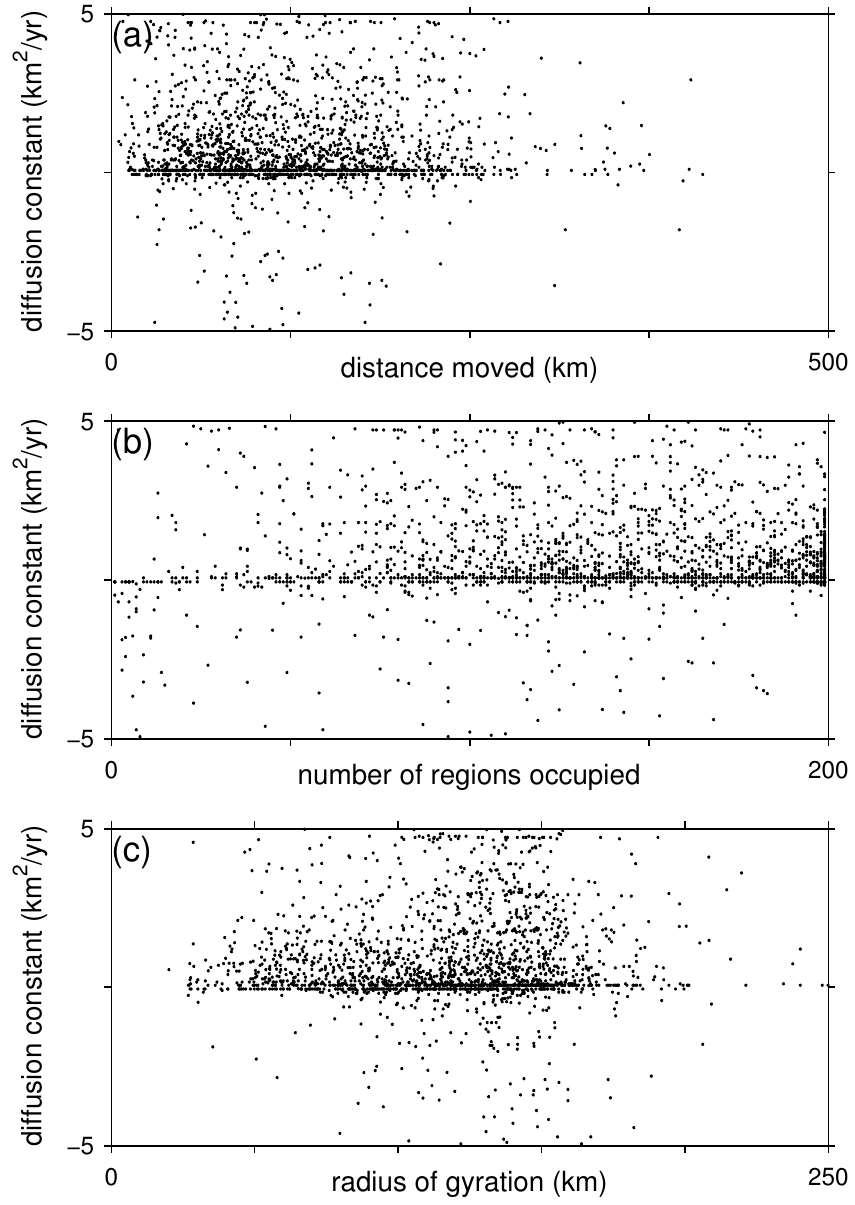} 
  \caption{Scatter plots of diffusion constants that we compute using the 1985 and 2000 census data versus (a) distance between origin and population centroid, (b) number of administrative regions, and (c) radius of gyration of clans in 2000. (In this figure, we consider the $3\,481$ clans for which data are available in both the 1985 and 2000 censuses; we were able to determine the origin location for $3\,120$ of these clans.) The corresponding Pearson correlation values are (a) $r \approx -0.02$ (from the $3\,120$ clans for which we know the origin location; the $p$-value is $p \approx 0.21$),
(b) $r \approx 0.088$ (from the $3\,481$ clans; the $p$-value is $p \approx 1.7 \times 10^{-7}$), and (c) $r \approx 0.097$ (from the $3\,120$ clans; the $p$-value is $p \approx 1.0 \times 10^{-8}$).
}
  \label{diffusion_distance_num_admin_RG}
\end{figure}

\begin{figure*}[b]
  \begin{tabular}{ll}
  \includegraphics[width=0.45\textwidth]{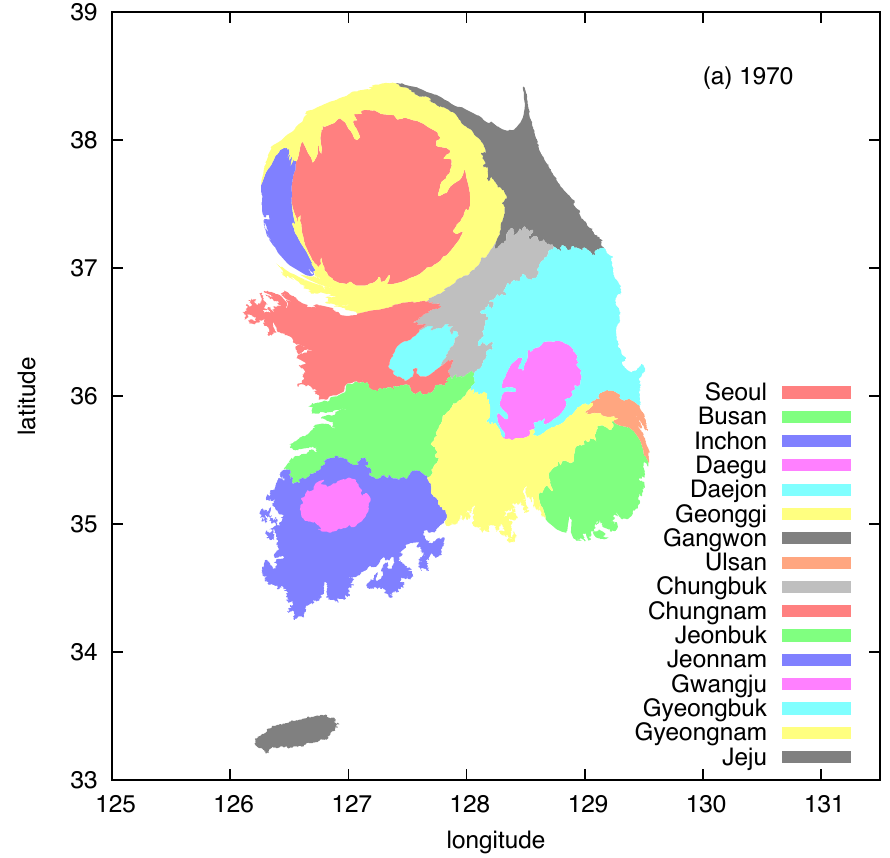} &
  \includegraphics[width=0.45\textwidth]{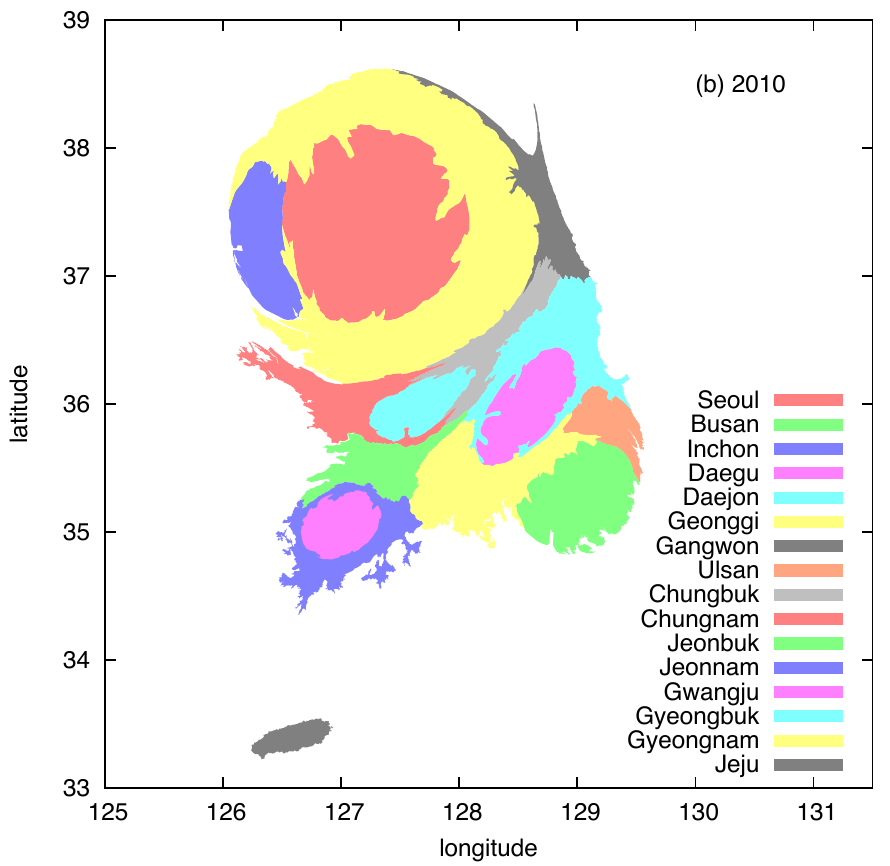} \\
  \end{tabular}
  \caption{Density-equalizing population cartograms~\cite{Gastner2004} for South Korea using population data from (a) 1970 and (b) 2010 censuses~\cite{population_census}. The coordinates are
longitude on the horizontal axis and latitude on the vertical axis.
The growth of the Seoul metropolitan area over the past 40 years is clearly visible. (Compare this figure to a regular map of South Korea, such as the one in Fig.~\ref{GimhaeKim_vs_HakseongLee} in the main text.)
}
  \label{cartogram}
\end{figure*}

\begin{figure}[t]
\includegraphics[width=0.9\columnwidth]{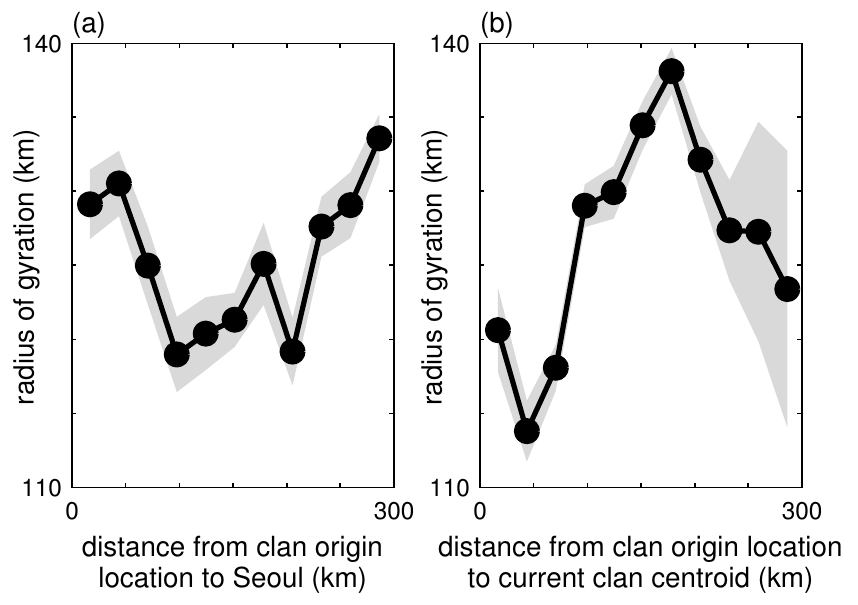} \\
\caption{
Radii of gyration and distance scales of of clans. (For this figure, we use the $3\,120$ clans that are present in both the 1985 and 2000 censuses and for which we could determine the origin location.)
(a) Radius of gyration $r_g$ versus distance to Seoul. 
The Pearson correlation between the variables is not statistically significant ($r \approx 0.18$; the $p$-value is $p \approx 0.6$).
(b) Radius of gyration $r_g$ versus distance between the clan origin location and the present-day centroid. The Pearson correlation between the diagnostics is positive and statistically significant up to 170 km ($r \approx 0.86$, $p \approx 0.01$) and is negative and significant for larger distances ($r \approx -0.96$, $p \approx 0.005$).
For each of the panels, we estimate $r_g$ separately in each of 11 equally-sized bins for the displayed range of distances. The gray regions give 95\% confidence intervals.
}
  \label{r_g_vs_dist_fig}
\end{figure}

\begin{figure}[t]
\includegraphics[width=0.9\columnwidth]{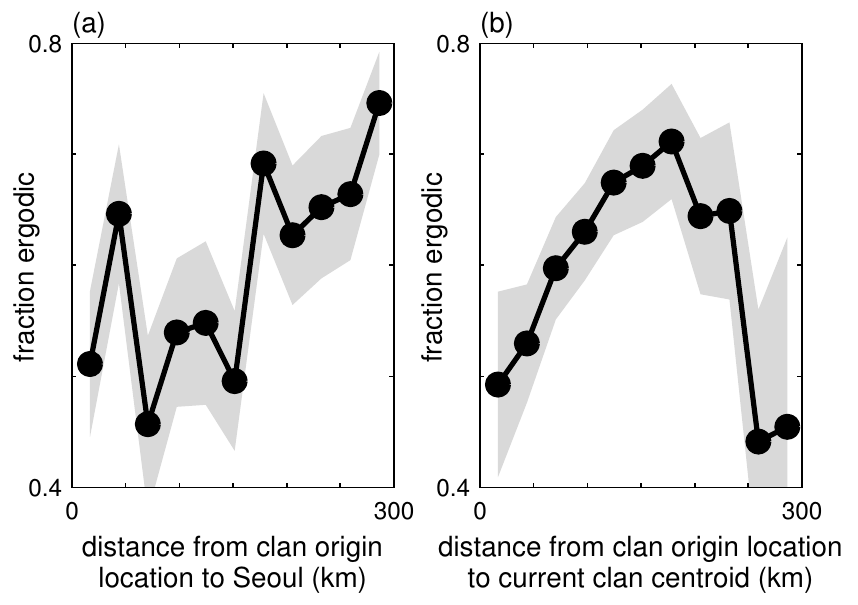} 
\caption{
Fraction of ergodic clans and 
distance scales of clans using only the $3\,120$ clans that we employed for the calculations in Fig.~\ref{r_g_vs_dist_fig}.  We obtain the same qualitative result as in Fig.~\ref{erg_vs_dist_fig} in the main text.
 (a) Fraction of ergodic clans versus distance to Seoul.  The correlation between the variables is positive and statistically significant. (The Pearson correlation coefficient is $r \approx 0.70$, and the $p$-value is $p \approx 0.02$.)  For the purpose of this calculation, we call a clan ``ergodic'' if it is present in at least 150 administrative regions.  (b) Fraction of ergodic clans versus the distance between the location of clan origin and the present-day centroid.  We measure ergodicity as in the left panel, and we estimate the fraction separately for each range of binned distances. (We use the same bins as in the left panel.) The correlation between the variables is positive and significant up to 170 km ($r \approx 0.99$, $p \approx 0.0001$) and is negative and significant for larger distances ($r \approx -0.92$, $p \approx 0.01$). 
For each of the panels, we estimate the fraction of ergodic clans in each of 11 equally-sized bins for the displayed range of distances. The gray regions give 95\% confidence intervals.
}
  \label{erg_vs_dist_fig_reduced_set}
\end{figure}

In this section, we discuss several figures that illustrate additional results. 
Figures \ref{phi_distribution_for_Kim_and_Lee}--\ref{diffusion_distance_num_admin_RG} explore clan ergodicity in more detail, and Fig.~\ref{cartogram} illustrates the ``convective'' effect of movement into the Seoul metropolitan area. Figures~\ref{r_g_vs_dist_fig} and \ref{erg_vs_dist_fig_reduced_set} provide alternative calculations from Fig.~\ref{erg_vs_dist_fig} in the main text.

In Fig.~\ref{phi_distribution_for_Kim_and_Lee}, we show the distribution of clan-density anomalies for the clans of the two Korean authors of this publication.  As we illustrated in Fig.~\ref{GimhaeKim_vs_HakseongLee}, Kim from Gimhae appears to be ergodic, whereas Lee from Hakseong appears to be localized.

In Fig.~\ref{distance_num_admin_RG}, we examine the correlation between the distance that a clan has moved and its current ergodicity.  We consider two measures of ergodicity---radius of gyration and number of regions occupied---and we also show the correlation between these two diagnostics for the clans. Some clans do not exist in the 2000 census data, and other clans only exist in one administrative region in 2000.  in 2000. As we indicate in Figs.~\ref{distance_num_admin_RG} and \ref{diffusion_distance_num_admin_RG}, we were able to determine the origin location for only $3\,120$ of the $3\,481$ clans that are in both the 1985 and 2000 census data, so several subsequent calculations only involve these 3120 clans. 
In Fig.~\ref{occu_num_diff_admin_Czech}, we show the distribution of clan ergodicities---using both number of regions occupied and radius of gyration---for the Czech Republic.  This is like Fig.~\ref{occu_num_diff_admin}, in which we showed this information for Korea. 
In Fig.~\ref{diffusion_distance_num_admin_RG}, we use scatter plots to examine the possible correlation between the calculated diffusion constants and the distance a clan has moved.  We similarly illustrate the connection between the diffusion constants and the two measures of ergodicity.

In Fig.~\ref{cartogram}, we show two ``cartograms'' \cite{Gastner2004} of South Korea.  In these images, we distort the administrative regions in proportion to the population of people who live there.  The growth of the Seoul metropolitan area over the past 40 years is clearly visible.

To examine an alternative characterization of ergodicity as the fraction of ergodic clans (see Fig.~\ref{erg_vs_dist_fig} in the main text), we examine radii of gyration $r_g$ versus distance to Seoul and versus distance between clan origin location and the present-day centroid.  We show our results in Fig.~\ref{r_g_vs_dist_fig}, and we see that they are qualitatively similar to those in Fig.~\ref{erg_vs_dist_fig}. For Fig.~\ref{r_g_vs_dist_fig}, we use the $3\,120$ clans that appear in both the 1985 and 2000 census data and for which we could determine the origin location. (Recall from Appendix~\ref{sec:geographical_information_from_google_maps} that we were able to identify the origin location of $3\,900$ of the $4\,303$ clans in the 2000 census data, and we used those $3\,900$ clans in Fig.~\ref{erg_vs_dist_fig} of the main text.) Consequently, we repeat the computation from Fig.~\ref{erg_vs_dist_fig} using this smaller set of clans. As one can see in Fig.~\ref{erg_vs_dist_fig_reduced_set}, we obtain the same qualitative result. (For this calculation, we use the clan centroids from the 2000 census.)


\end{CJK}  


\begin{thebibliography}{00}
\bibitem{jokbo_details} An original \jokbo{} usually also includes other details about clans, but we only use information that we mention explicitly.
\bibitem{southKoreaabund} In fact, it is known that clans that originated in the southern part of Korea (modern-day South Korea) have been more abundant than the northern part (modern-day North Korea) in recent Korean history (including the period that is spanned by our data set)~\cite{KoreanFamilyNamesAndGenealogies}.
\bibitem{ten_possibilities} Because we only have ten family books, our data allow only ten distinct possibilities for ``clan $j$'' (the family into which a woman marries).
\bibitem{origin_locations} We did attempt to determine clan origin locations; see Fig.~\ref{movement_to_cm_fig}.
\bibitem{Thornthwaites_law} One might call this ``Thornthwaite's law'' of human diffusion.
\bibitem{manipulation} In practice, we manipulate the anomaly data before computing radii of gyration. See Appendix~\ref{sec:estimating_diffusion_constants} for details.
\end{thebibliography}
\end{document}